\documentclass[trackchanges,twocolumn]{aastex701}
\usepackage{amsmath,amssymb}
\usepackage{multirow}
\usepackage{float}
\usepackage[version=4]{mhchem} 
\usepackage{tabularx}
\usepackage{threeparttable}
\usepackage{booktabs}
\usepackage{siunitx}
\usepackage{subfigure}
\usepackage{comment}
\usepackage{tablefootnote}
\usepackage{CJK}

\begin{document}
\begin{CJK*}{UTF8}{gbsn}
\title{Does Topology-Dependent Chemistry Influence the Interstellar PAH Inventory? Insights from the Absence of Cyanotriphenylenes}

\author[0009-0003-9416-5550]{Huanyu Ren(任焕宇)}
\affiliation{Department of Chemistry, Massachusetts Institute of Technology, Cambridge, MA 02139, USA}
\email{hyren@mit.edu}

\author[0000-0002-0332-2641]{Gabi Wenzel}
\affiliation{Department of Chemistry, Massachusetts Institute of Technology, Cambridge, MA 02139, USA}
\affiliation{Center for Astrophysics \textbar{} Harvard \& Smithsonian, Cambridge, MA 02138, USA}
\email{gwenzel@mit.edu}

\author[0000-0001-8134-5681]{Thomas H. Speak}  
\affiliation{Department of Chemistry, University of British Columbia, 2036 Main Mall, Vancouver BC V6T 1Z1, Canada}
\email{tspeak@chem.ubc.ca}

\author[0009-0007-2048-2907]{Miya Duffy}
\affiliation{Department of Chemistry, Massachusetts Institute of Technology, Cambridge, MA 02139, USA}
\email{mfduffy@mit.edu}

\author[0000-0002-9995-2700]{Annapoorani Hariharan}
\affiliation{Department of Chemistry, Massachusetts Institute of Technology, Cambridge, MA 02139, USA}
\email{annakh15@mit.edu}

\author[0009-0008-1171-278X]{Martin S. Holdren}
\affiliation{Department of Chemistry, Massachusetts Institute of Technology, Cambridge, MA 02139, USA}
\email{holdrenm@mit.edu}

\author[0009-0005-1773-8460]{D. Archie Stewart} 
\affiliation{Department of Chemistry, Massachusetts Institute of Technology, Cambridge, MA 02139, USA}
\email{archie25@mit.edu}

\author[0000-0002-5171-7568]{Christopher N. Shingledecker}
\affiliation{Department of Chemistry, Virginia Military Institute, Lexington, VA 24450, USA}
\email{shingledeckercn@vmi.edu}

\author[0009-0002-6372-9926]{Reace H. J. Willis}  
\affiliation{Department of Chemistry, University of British Columbia, 2036 Main Mall, Vancouver BC V6T 1Z1, Canada}
\email{rwillis@chem.ubc.ca}

\author[0000-0002-0850-7426]{Ilsa R. Cooke}  
\affiliation{Department of Chemistry, University of British Columbia, 2036 Main Mall, Vancouver BC V6T 1Z1, Canada}
\email{icooke@chem.ubc.ca}

\author[0000-0003-1254-4817]{Brett A. McGuire}
\affiliation{Department of Chemistry, Massachusetts Institute of Technology, Cambridge, MA 02139, USA}
\affiliation{National Radio Astronomy Observatory, Charlottesville, VA 22903, USA}
\email{brettmc@mit.edu}

\correspondingauthor{Huanyu Ren, Brett A. McGuire}
\email{hyren@mit.edu, brettmc@mit.edu}

\begin{abstract}
Although a set of polycyclic aromatic hydrocarbons (PAHs) has recently been identified in the cold dark cloud TMC-1, several common PAH motifs remain absent, leaving the origin of this apparent structural selectivity unresolved. Triphenylene (\ce{C18H12}) provides a stringent test of whether relative thermodynamic stability alone can account for the emerging PAH inventory because it is the smallest fully benzenoid PAH predicted by Clar's rule, is exceptionally stable, and is a key subunit of larger detected aromatic frameworks such as coronene and fullerene-related species. To probe whether such thermodynamically favored carbon skeletons are selected under cold interstellar conditions, we synthesized cyanotriphenylenes and measured their first 6.5--16~GHz microwave rotational spectra as polar tracers for the non-polar parent. An astronomical search based on these spectroscopic parameters yielded no detection, suggesting a possible ``missing triphenylene'' anomaly: despite its exceptional thermodynamic stability, this fully benzenoid PAH may be relatively depleted in TMC-1. To rationalize this behavior, we introduce a ring-node graphic representation that classifies PAH skeletons by topology; within this framework, triphenylene, whose compactness is otherwise ambiguous, is classified as non-compact. This representation reveals a tentative clustering of the currently detected and undetected PAHs according to topology, suggesting that the current PAH inventory cannot be rationalized solely on the basis of thermodynamic stability and may instead reflect differences in formation and destruction kinetics. The observed selectivity may arise from two complementary effects: sequential growth decay along bottom-up growth pathways and rapid consumption of bay-region intermediates by reactive unsaturated small species.
\end{abstract}

\keywords{\uat{Astrochemistry}{75}; \uat{Interstellar medium}{847}; \uat{Molecular clouds}{1072}; \uat{Microwave spectroscopy}{2251}; \uat{Rotational spectroscopy}{2248}; \uat{Polycyclic aromatic hydrocarbons}{1280}; \uat{Interstellar molecules}{849}; \uat{Radio astronomy}{1338}; \uat{Chirped-pulse Fourier-transform}{2254}}
\section{Introduction}
Polycyclic aromatic hydrocarbons (PAHs) are a family of carbon-containing, highly unsaturated molecules that are often formed through combustion processes because of their special thermodynamic stability. Their stability also makes these molecules favorable for survival in extraterrestrial environments, leading to great interest in searching for them in the interstellar medium (ISM), such as in the Taurus Molecular Cloud (TMC-1). Roughly 10--25\% of the cosmic carbon in the diffuse ISM is thought to be locked in PAHs \citep{PAHsincarbon,PAHsincarbon2,BrettMcCarthyReview}, making them central to interstellar carbon chemistry. Beyond expanding the molecular census, the wide structural diversity of PAHs makes them a useful platform for probing assembly and selectivity of large aromatic molecules under extreme interstellar conditions.

\begin{figure}[htb!]
    \centering
    \includegraphics[width=\linewidth]{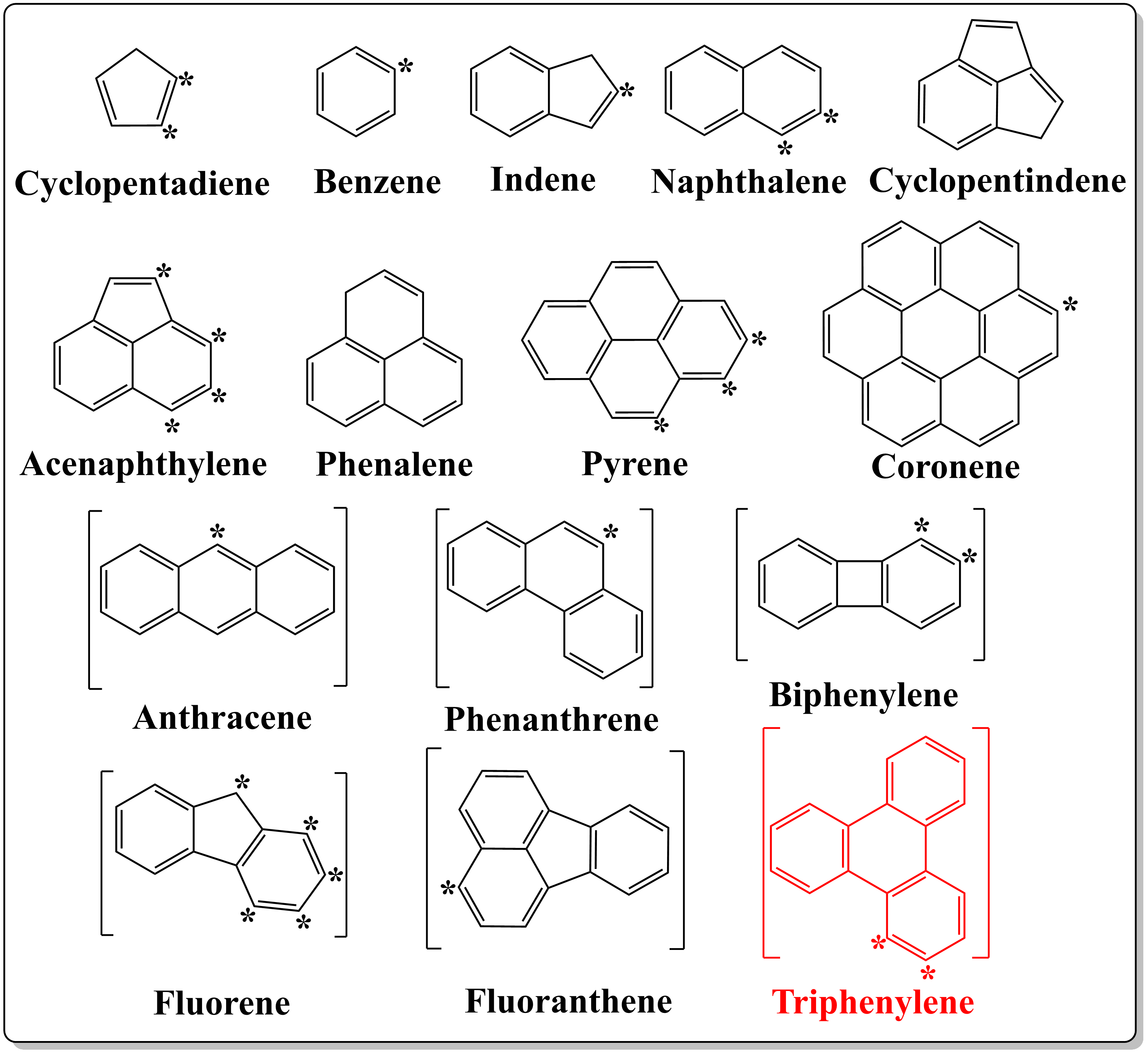}
    \caption{Overview of cyclic carbon skeletons with published spectroscopic data relevant to this work, while asterisks mark the positions of the cyano group for nitrile derivatives, including species detected in TMC-1 and representative PAHs that remain undetected despite the availability of laboratory microwave data for either the parent molecule or nitrile derivatives. The spectroscopy and non-detection of cyanotriphenylenes, marked with red color, are reported in this work. Species enclosed in square brackets denote non-detections in astronomical searches.}
    \label{fig:intro}
\end{figure}

Recent detections show that the interstellar PAH inventory is broadening in structural complexity. The unsubstituted hydrocarbons 1H-indene, 1H-cyclopent[cd]indene, and 1H-phenalene have been detected directly in TMC-1 \citep{indeneB,indeneC,cyclopentindene,phenalene}. In addition, the naphthalene, acenaphthylene, pyrene, and coronene carbon skeletons have been identified indirectly through detections of their polar cyano derivatives \citep{cyanonaphthalene,cyanoacenaphthylene,cyanoacenaphthylene2,cyanopyrene,cyanopyrene2,cyanocoronene}; 2-cyanoindene has likewise been detected alongside the parent indene \citep{cyanoindene}. This expanding molecular inventory has been enabled by deep low-frequency microwave surveys of TMC-1, particularly the 100~m Green Bank Telescope Observations of TMC-1: Hunting Aromatic Molecules (GOTHAM) program \citep{GOTHAM} and the Yebes 40~m telescope Q-band Ultrasensitive Inspection Journey to the Obscure TMC-1 Environment (QUIJOTE) program \citep{benzyne}. At the same time, many common PAHs or their cyano derivatives (with 15--20 carbon atoms) that are accessible in the laboratory remain undetected astronomically, including biphenylene \citep{cyanobiphenylene}, fluorene \citep{Fluorene, cyanofluorene, iminofluorene}, anthracene \citep{cyanophenanthrene}, phenanthrene \citep{Fluorene, cyanophenanthrene}, and fluoranthene \citep{cyanofluoranthene} as summarized in Fig.~\ref{fig:intro}. The emerging contrast is therefore not simply one of molecular size, but of which PAH topologies are preferentially retained in TMC-1, pointing to chemically informative structural selectivity in interstellar PAH formation and survivability.

\begin{figure}[htb!]
    \centering
    \includegraphics[width=0.7\linewidth]{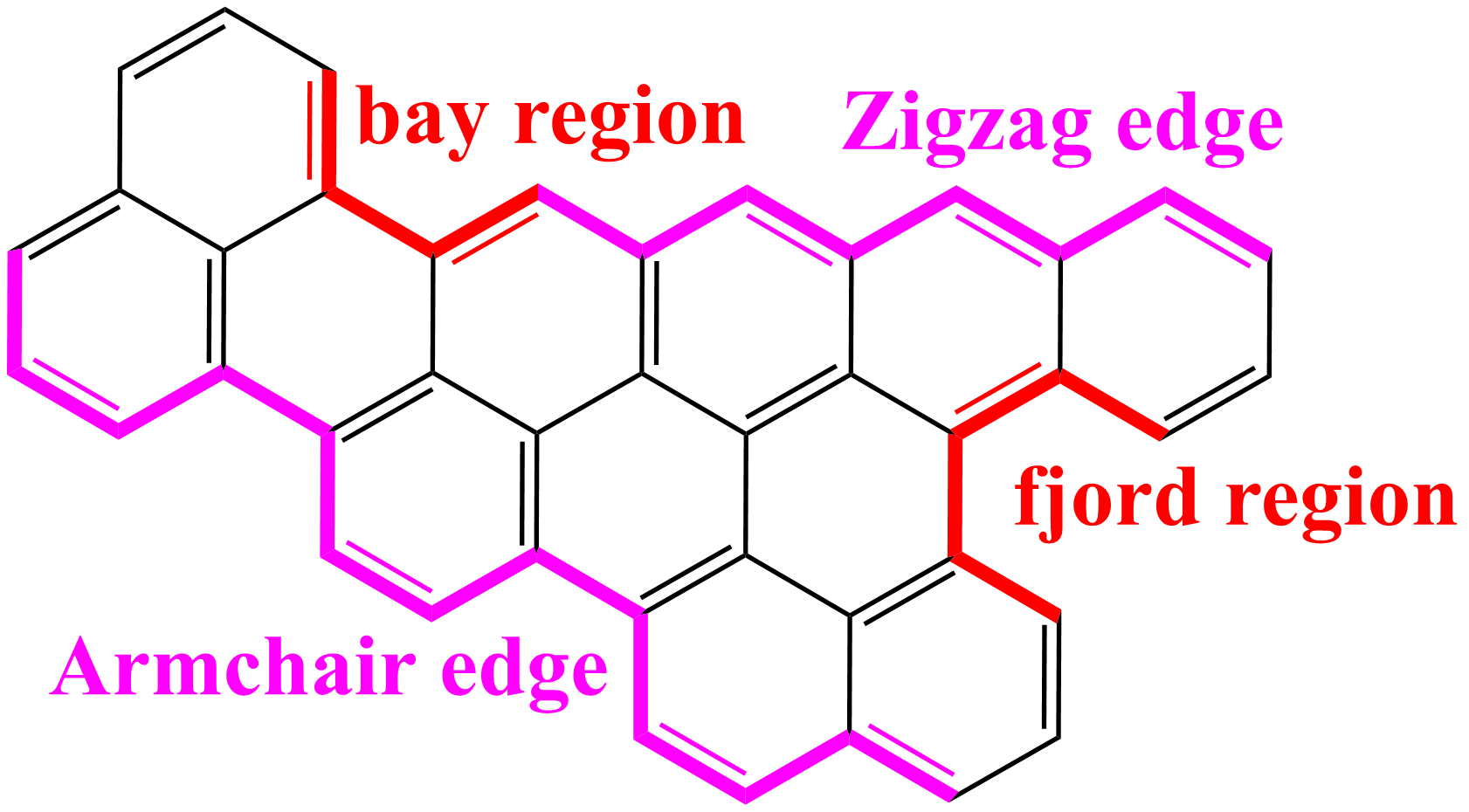}
    \caption{Schematic definitions of key PAH edge motifs and local topological features used in the structural discussion. Zigzag and armchair describe global edge motifs, whereas bay and fjord denote local edge topologies. \citep{Ricca2019}}
    \label{fig:edge}
    
\end{figure}
Such structural selectivity has long been inferred from the infrared signatures of interstellar PAH ensembles, with several of the relevant edge motifs illustrated in Fig.~\ref{fig:edge}. Early analyses of the 10--15~$\mu$m CH out-of-plane bands established that variations across astronomical sources reflect changes in hydrogen adjacency classes and, by extension, in PAH edge structure \citep{Hony2001}. In a complementary spectral region, the 6--9~$\mu$m PAH emission bands, dominated by C--C stretching and C--H in-plane bending motions, were classified into Class A/B/C profiles, showing that distinct astrophysical environments host systematically different PAH sub-populations, including ionized PAHs \citep{Peeters2002}. Variations in the aromatic C--H stretching band at 3.3~$\mu$m and the aromatic C--H out-of-plane bending band at 11.2~$\mu$m were later likewise resolved into spectral subclasses \citep{vanDiedenhoven2004}, reinforcing the view that interstellar PAH populations exhibit environment-dependent structural biases. Subsequent modeling associated Class B-type emission with armchair-edge, more bay-containing PAH substructures \citep{Ricca2019}, whereas recent James Webb Space Telescope (JWST) observations of the photodissociation region (PDR) at the Orion Bar, a region dominated by Class A-like PAH emission \citep{Chown2024}, were interpreted in terms of a more compact, zigzag-rich PAH family with only minor bay/fjord contributions under strong UV processing \citep{orionbarJWSTPAH}.

However, although mid-infrared PAH emission studies yield rich structural constraints on interstellar PAH ensembles, such observations are inherently limited to warm, strongly irradiated environments in which the molecular population has already been filtered by photoprocessing \citep{Photoprocessing1,Photoprocessing2}. Microwave rotational spectroscopy offers the complementary advantage of isomer-specific molecular identification in cold, weakly irradiated sources such as TMC-1, where PAHs are otherwise inaccessible through vibrational emission. With the growing number of structurally distinct PAHs now identified in TMC-1, this approach is beginning to reveal whether cold interstellar PAH chemistry follows a characteristic topology-dependent selection.

\begin{figure}[htb!]
    \centering
    \includegraphics[width=\linewidth]{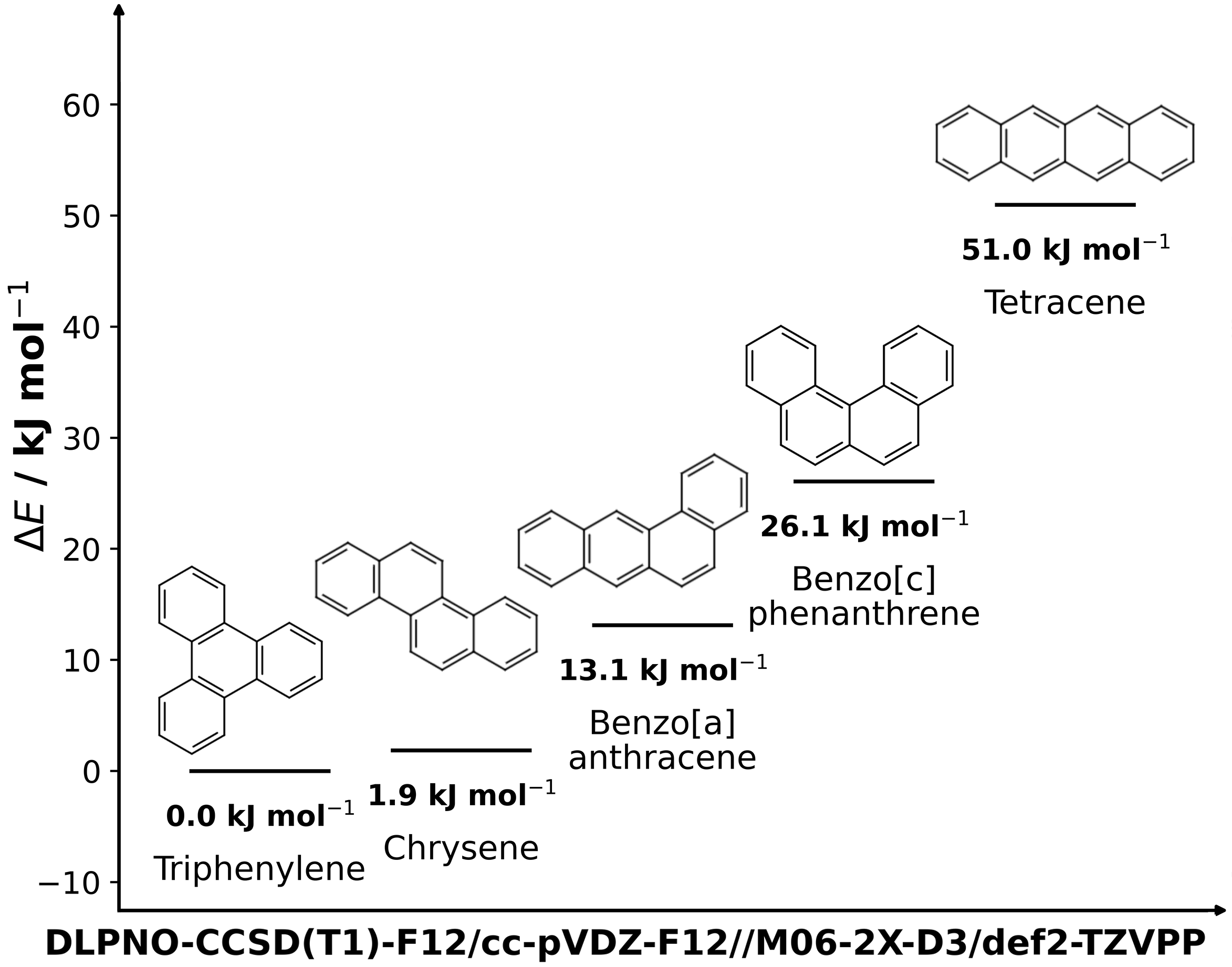}
    \caption{Quantum mechanical calculations at the DLPNO-CCSD(T1)-F12/cc-pVDZ-F12//M06-2X-D3/def2-TZVPP level of theory indicate that triphenylene is the most stable isomer in the gas phase within the unsubstituted benzenoid \ce{C18H12} isomer family.}
    \label{fig:energies}
\end{figure}

As summarized in Fig.~\ref{fig:intro}, the PAH skeletons detected in TMC-1 already suggest a clear structural clustering toward compact frameworks, with little representation of bay- or fjord-containing topologies. However, the descriptors implicated by this trend are not fully equivalent: overall compactness and edge topology need not coincide. Triphenylene is therefore a particularly revealing structural test case. Although its carbon skeleton is sometimes considered compact and highly benzenoid, its bay-containing framework is also the smallest PAH composed entirely of armchair edges \citep{Ricca2019}. It thus occupies a boundary position relative to previously searched PAHs, combining a relatively compact carbon skeleton with armchair/bay-related edge character.

A separate but central question is whether such structural selectivity in TMC-1 is driven primarily by thermodynamic stability or by kinetic constraints on PAH formation and processing. Within the four-ring \ce{C18H12} isomer family, triphenylene is the thermodynamic minimum, consistent with Clar's rule, which relates PAH stability to the maximization of disjoint aromatic $\pi$-sextets (benzene-like moieties), and with its exceptional chemical inertness \citep{inert,Clar}. The relative energies shown in Fig.~\ref{fig:energies} further establish triphenylene as the most stable member of this isomeric manifold. Clar's rule therefore provides a useful thermodynamic prior for assessing PAH stability and survivability \citep{BrettMcCarthyReview}. Triphenylene consequently represents a near-maximal test of thermodynamic selection: if the Clar-favored \ce{C18H12} minimum is absent while less thermodynamically privileged PAH frameworks are present, then thermodynamic stability alone cannot account for the PAH inventory in TMC-1. Instead, such an outcome would point to kinetic or topological constraints as key determinants of which PAH skeletons are formed or preserved in this source, as has been seen in other interstellar isomer families: the lowest-energy \ce{H2C3O} isomer propadienone long remained undetected despite detections of propynal and cyclopropenone from Sgr B2(N) and TMC-1 \citep{H2C3O1, H2C3O2, H2C3O3}.

Triphenylene is not merely a structural and thermodynamic test case, but also a chemically plausible interstellar target. It has not only been regarded as a potential diffuse interstellar band (DIB) carrier \citep{TriphenyleneDIBs}, but has also been identified in primitive extraterrestrial materials, including the Murchison meteorite \citep{meteorite} and samples returned from asteroids Ryugu \citep{Ryugu} and Bennu \citep{Bennu}, showing that this fully benzenoid framework can survive in carbon-rich interstellar condensed states. 

This plausibility may extend beyond solid-phase survival. A previous combined experimental and theoretical study proposed an unconventional hydrogen abstraction-vinylacetylene addition (HAVA) pathway as a possible gas-phase route to triphenylene under cold astrophysical conditions \citep{gasphase}. The experiment reported a product signal at the appropriate mass-to-charge ratio, while the calculated potential-energy surface suggested that triphenylene could be one of the products. However, the assignment was not isomer-specific, product branching ratios were not determined, and the low-temperature viability of the pathway may depend sensitively on a shallow entrance-barrier region. This work therefore provides a plausible, but not definitive, route to triphenylene formation in cold gas-phase environments.

Triphenylene is also structurally relevant to larger interstellar aromatic frameworks. It is a benzenoid subunit of both coronene and fullerene-related carbon skeletons. Although fullerenes have not been identified in TMC-1, \ce{C60} has been widely detected in other astrophysical environments \citep{C60}, and the ion \ce{C60+} has been assigned as a carrier of several DIBs \citep{C60+1,C60+2,C60+3}. More directly relevant to TMC-1, coronene has already been indirectly detected there through observation of its cyano derivative, cyanocoronene \citep{cyanocoronene}. Because triphenylene can be viewed as a smaller structural fragment of coronene, searching for it may also provide insight into how larger compact aromatic frameworks emerge in cold interstellar environments.

\begin{figure}[htb!]
    \centering
    \includegraphics[width=\linewidth]{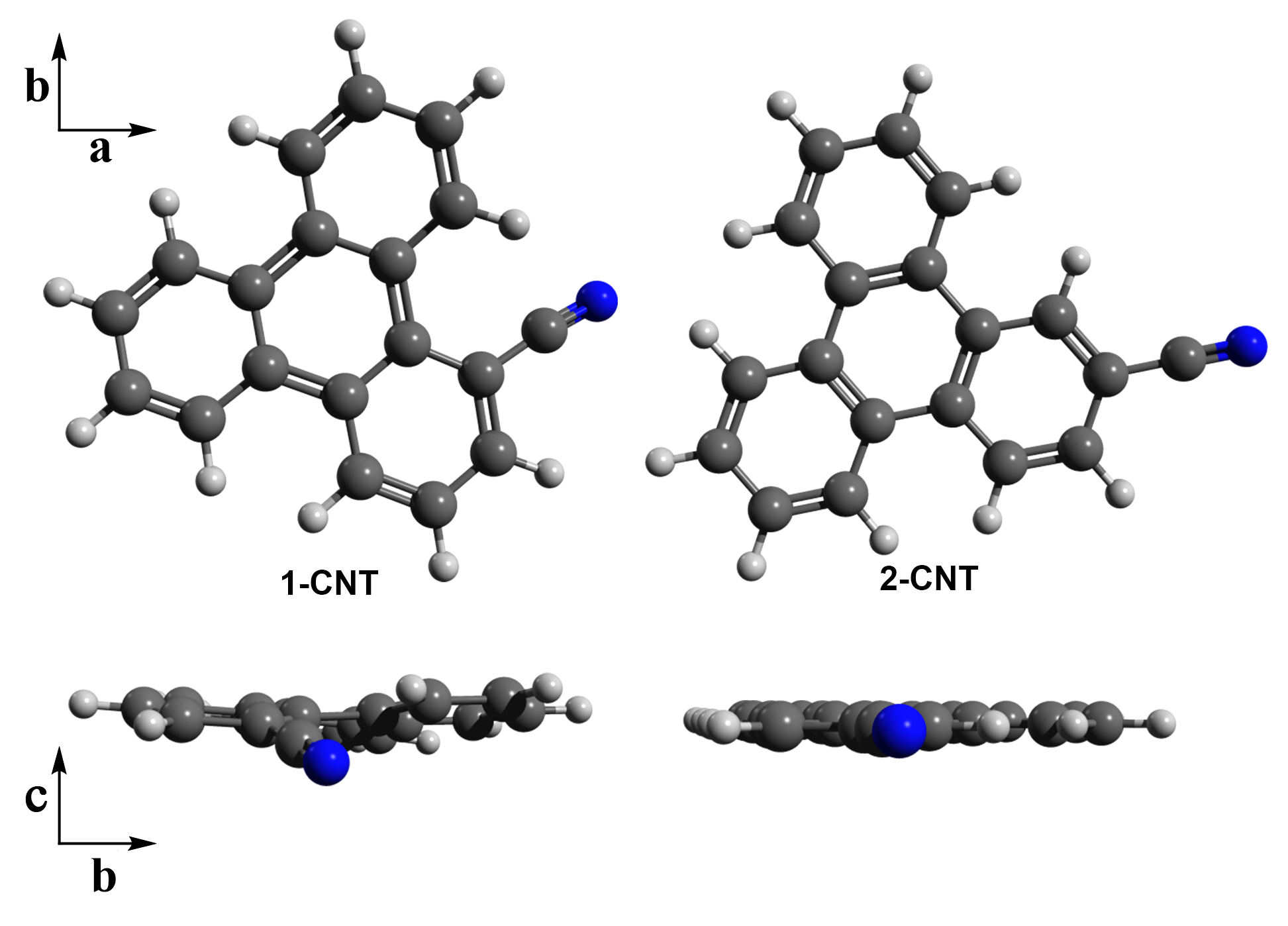}
    \caption{The B3LYP-D3(BJ)/6-311++g** level optimized geometry of 1-cyanotriphenylene (1-CNT) and 2-cyanotriphenylene (2-CNT). 2-CNT has a fully planar structure, while 1-CNT is non-planar with a permanent electric dipole moment along the $c$-axis, caused by the steric hindrance of the bay-region cyano group.}
    \label{fig:molecule}
\end{figure}

To search for non-polar molecules like triphenylene with rotational spectroscopy, adding a polar group such as \ce{CN} is a widely applied strategy for the indirect detection of symmetric PAHs, given the broadly barrierless cyano-addition reactions of PAHs \citep{CN+bezene,cyanonaphthalene,cyanopyrene2,cyanocoronene}. However, when many cyano-substituted isomers are possible, the abundance of the cyano-PAH population would be partitioned among several structures, reducing the column density and astronomical line intensity of each individual isomer. In the case of triphenylene, cyanation yields only two isomers: 1-cyanotriphenylene and 2-cyanotriphenylene (1-CNT and 2-CNT, respectively; their structures are shown in Fig.~\ref{fig:molecule}). In this work, we synthesized and measured the rotational spectra of both 1-CNT and 2-CNT to search for them in TMC-1.

\section{Methods}
\subsection{Microwave Spectroscopy}\label{subsec:spectroscopy}

1-CNT and 2-CNT were synthesized from bromotriphenylenes by the Rosenmund-von Braun reaction and purified by flash column chromatography \citep{cyanopyrene2}. Details of the synthesis are provided in the Supporting Information.

The broadband rotational spectra of 1-CNT and 2-CNT (from 6.5 to 16 GHz) were measured using a chirped-pulse Fourier transform microwave (CP-FTMW) spectrometer \citep{Tweety}. The sample was heated to 180~\textdegree C to achieve detectable vapor pressures. Neon was chosen as the carrier gas with a backing pressure of 1.28 kTorr, which mixed with the sample vapor, and subsequently was pulsed into the $10^{-5}$ Torr vacuum chamber with a mass-flow rate of 3.8 sccm. The gas pulses were generated by a commercial Parker Series 9 solenoid valve with 3~Hz repetition frequency, held open for a duration of 950 $\mu$s. After a 1200 $\mu$s delay, 650 ns long microwave chirp pulses were sent to excite molecules. The 20 $\mu$s free induction decays (FIDs) were then recorded for higher spectroscopic resolution of hyperfine splitting. Such an excitation-FID collection cycle was repeated 10 times for each gas pulse. To achieve sufficient signal-to-noise ratios for analysis, five million FIDs were integrated in the time domain, filtered for instrumental artifacts, filtered with a Hann window function, zero-padded, and finally Fourier transformed into the frequency domain as the target spectrum.

\subsection{Quantum Chemical Calculations}\label{subsec:computing}
To predict rotational constants for the spectral fitting procedure, quantum chemical calculations were performed with the Gaussian 16 suite of programs (G16.C.02 version, \citealt{g16c02}) using the B3LYP-D3(BJ)/6-311++G(d,p) method and basis set \citep{B3LYP,6-311g,D3BJ} for geometry and harmonic vibrational frequency analysis. The relative energy calculations of \ce{C18H12} isomers were performed with the DLPNO-CCSD(T1)-F12/cc-pVDZ-F12//M06-2X-D3/def2-TZVPP method and basis set. Geometry optimization and the zero-point energies (ZPEs) of isomers were first computed with the M06-2X functional \citep{Zhao:2008:1432} with the D3(0) empirical dispersion correction \citep{D3} and the triple zeta def2-TZVPP basis set \citep{def2TZVPP1,def2TZVPP2} via Gaussian 16. Then, high-level single-point electronic energies (SPEs) were calculated with the domain-based local pair natural orbital approach to coupled cluster with single and double excitations with perturbative triples (DLPNO-CCSD(T1)-F12, \citealp{Pinski:2015:034108,Riplinger:2016:024109} tight PNO thresholds), and applied cc-pVDZ-F12 as main basis set and complementary auxiliary basis sets (CABS), cc-pVTZ as auxiliary basis sets, via the ORCA 6.1.0 program \citep{orca2020, ORCA:2025}. The detailed ab initio calculation results for the potential energy surface of cyano-addition reactions are available in the Supporting Information.

\begin{figure*}[t!]
\begin{center}
\includegraphics[width=\textwidth]{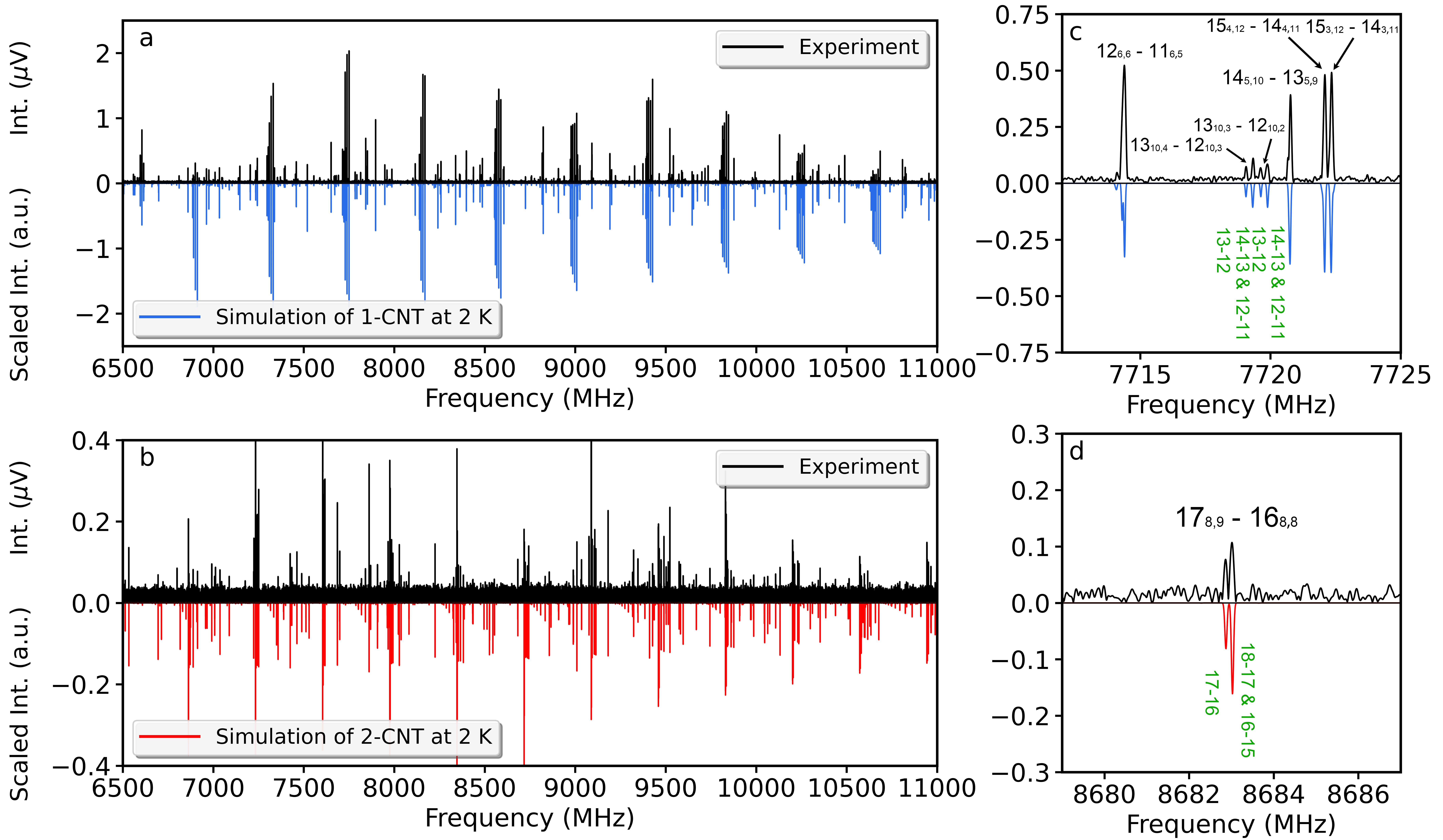}    
\caption{Experimental and simulated rotational spectra of 1-CNT and 2-CNT. The simulated spectra were convolved with a Gaussian line profile with a full width at half maximum (FWHM) of 0.06 MHz. Panels a and b show a part of experimental spectra from 6.5 to 11~GHz compared with the simulated stick spectra at 2 K, while panels c and d present expanded views of selected spectral regions to show some partially resolved hyperfine splitting.}
\label{fig:spectra}
\end{center}
\end{figure*}

\begin{table*}[t]
\centering
\begin{threeparttable}
\caption{Rotational and hyperfine constants of 1-cyanotriphenylene and 2-cyanotriphenylene experimentally fit in the A-reduced I$^r$ representation compared with the ones resulting from calculations at the B3LYP-D3(BJ)/6-311++G(d,p) level of theory.}
\label{tab:constants}
\footnotesize
\begin{tabular}{l rr rr}
\toprule
Parameter & \multicolumn{2}{c}{1-CNT} & \multicolumn{2}{c}{2-CNT} \\
\cmidrule(lr){2-3} \cmidrule(lr){4-5}
 & Computational & Experimental\tnote{a} & Computational & Experimental\tnote{a} \\
\midrule
$A$ (MHz)                            & 502.11   & 501.470180(93)   & 513.35   & 512.60840(39)   \\
$B$ (MHz)                            & 352.64   & 351.994684(61)   & 290.85   & 290.419960(40)   \\
$C$ (MHz)                            & 210.13   & 209.631627(39)   & 185.66   & 185.483694(50)  \\
$\Delta_J$ (Hz)                      & 2.01     & 2.386(90)        & 0.71     & 0.810(28)        \\
$\Delta_{JK}$ (Hz)                   & $-$2.41  & $-$3.09(41)      & 4.04     & [4.04]\tnote{b}        \\
$\Delta_K$ (Hz)                      & 7.00     & 8.67(57)         & $-$0.87  & [$-$0.87]\tnote{b} \\
$\delta_J$ (Hz)                      & 0.41     & 0.510(43)        & 0.25     & [0.25]\tnote{b}      \\
$\delta_K$ (Hz)                      & 2.40     & 2.72(29)         & 3.03     & 3.89(28)        \\
$\frac{3}{2}\chi_{aa}$ (MHz)         & $-$3.79  & $-$3.687(17)     & $-$6.23  & $-$5.825(58)     \\
$\frac{1}{4}(\chi_{bb} - \chi_{cc})$ (MHz)
                                           & $-$0.222 & $-$0.2280(48)    & 0.048    & [0.048]\tnote{b} \\
\addlinespace
$|\mu_a|$ (D)                        & 4.3      & Observed         & 5.6      & Observed        \\
$|\mu_b|$ (D)                        & 1.5      & Observed         & 0.7      & Not Observed    \\
$|\mu_c|$ (D)                        & 1.0      & Not Observed     & 0.0      & Not Observed    \\
\addlinespace
$N^\mathrm{fit}_\mathrm{lines}$\tnote{c}
                                           & \textemdash & 374          & \textemdash & 253 \\
$N^\mathrm{hyperfine}_\mathrm{lines}$\tnote{d}
                                           & \textemdash & 59           & \textemdash & 40 \\
$\sigma_\mathrm{fit}$ (kHz)          & \textemdash & 11.4          & \textemdash & 11.9 \\
$\Delta$ (u \AA$^2$)\tnote{e}        & $-$34.58  & $-$32.7566(6)     & 0.00     & $-$1.4084(21)    \\
$(J, K_a)_\mathrm{max}$              & \textemdash & (39, 16)      & \textemdash & (40, 14) \\
\bottomrule
\end{tabular}

\begin{tablenotes}
    \item[a] Values in parentheses are 1$\sigma$ uncertainties in units of the last digit.
    \item[b] Values in square brackets are frozen from the computation because of the relatively large error in the fit.
    \item[c] The number of assigned experimental peaks. One peak can be assigned to several transitions due to limited resolution.
    \item[d] The number of hyperfine-resolved groups, in which transitions have the same $J$, $K_a$, and $K_c$ but different experimental frequencies.
    \item[e] Inertial defect, defined as $\Delta = I_c - I_a - I_b$, where $I_i = 505379.006/B_i$ for rotational constants in MHz and moments of inertia in u \AA$^2$.
\end{tablenotes}
\end{threeparttable}
\end{table*}

\section{Results and Discussion} \label{sec:results}
\subsection{Laboratory Measurements}\label{subsec:measurements}

Due to the high melting point of 2-CNT ($215$~\textdegree C--$216$~\textdegree C, reported by \citealt{NMR1}), even with the elevated temperature of the nozzle, the vapor pressure of 2-CNT in the supersonic expansion was limited. Consequently, the signal-to-noise ratio (SNR) of its spectrum is modest. Moreover, the permanent electric dipole moment component of 2-CNT lies almost entirely along the $a$-axis, and hence only $a$-type transitions were detectable with our achievable SNR. The absence of $b$-type transitions limits the sampling of transitions connecting different $K_a$ ladders and therefore weakens the independent constraints on both $A$ and the $K$-dependent centrifugal-distortion term $\Delta_K$. Together with the modest spectral SNR, this incomplete transition-type coverage produced the large fitted uncertainties in $\Delta_K$, $\Delta_{JK}$, $\delta_{J}$ and $\frac{1}{4}(\chi_{bb} - \chi_{cc})$; both constants were consequently fixed to their calculated values in the final fit of 2-CNT.

Table~\ref{tab:constants} provides all the least-squares fitted rotational constants, centrifugal distortion parameters, and $^{14}$N nuclear quadrupole hyperfine coupling constants determined using the SPCAT/SPFIT suite \citep{spcat} and formatted with PIFORM \citep{KisielPIFORM}, while the computational values are also listed for comparison. The $^{14}$N nuclear quadrupole interaction was included throughout the spectral simulation and fitting procedure, using the quantum-chemically calculated coupling constants as the initial values. For unresolved hyperfine blends, multiple transitions were assigned to a common experimental frequency in SPFIT and weighted according to their relative simulated intensities. When shoulders or multiple components were experimentally distinguishable, Gaussian line-shape fitting was used to obtain separate experimental center frequencies, which were then assigned to the corresponding groups of predicted hyperfine transitions. Ray's asymmetry parameter ($\kappa=[2B-A-C]/[A-C]$) was determined to be -0.024 for 1-CNT and -0.36 for 2-CNT, indicating that 2-CNT exhibits more pronounced prolate character, while 1-CNT lies close to the maximally asymmetric rotor limit. Based on this, the spectral analysis was performed using Watson's A-reduced Hamiltonian in the $I^r$ representation.

The planarity of the two isomers was evaluated using the inertial defect, which compares the moment of inertia about the axis perpendicular to the molecular plane with the sum of the two in-plane moments. The principal moments of inertia are related to the rotational constants, and the inertial defect can therefore be written as
\begin{equation}
\Delta = I_c - I_a - I_b
= \frac{h}{8\pi^2}
\left(
\frac{1}{C}
-
\frac{1}{A}
-
\frac{1}{B}
\right),
\label{eq:inertial_defect}
\end{equation}
where $A$, $B$, and $C$ are the rotational constants associated with the principal axes $a$, $b$, and $c$, respectively. In evaluating eq~\ref{eq:inertial_defect}, the rotational constants were converted to SI units, and the resulting moments of inertia were expressed in u~\AA$^2$.

For a perfectly planar rigid molecule, the perpendicular-axis theorem requires $I_c = I_a + I_b$, and therefore $\Delta = 0$. In practice, small deviations from zero can arise from zero-point vibrational averaging and other non-rigid effects. Using equation~\ref{eq:inertial_defect}, the experimentally determined inertial defect of 2-CNT is $-1.41$ u~\AA$^2$, consistent with an approximately planar heavy-atom framework. In contrast, 1-CNT gives a much larger negative inertial defect of $-32.76$ u~\AA$^2$, far beyond what would normally be expected from vibrational averaging of a planar PAH. This pronounced breakdown of the perpendicular-axis relationship provides direct spectroscopic evidence that 1-CNT adopts a substantially non-planar geometry, whereas 2-CNT remains close to planar, in accordance with the optimized geometry shown in Figure~\ref{fig:molecule}.

\subsection{Astronomical Search toward TMC-1}\label{subsec:searching}
The search for 1- and 2-cyanotriphenylene in TMC-1 utilized observational data from the 100~m Robert C. Byrd Green Bank Telescope (GBT) acquired as part of the GOTHAM large program \citep{GOTHAM}, specifically the fifth data reduction (DR5.3)\citep{DR5.3}. The GBT was pointing toward the ``cyanopolyyne peak'' in TMC-1 at right ascension $\alpha_\mathrm{J2000} = 04^{\mathrm{h}}41^{\mathrm{m}}42.5^{\mathrm{s}}$, and declination $\delta_\mathrm{J2000} = +25^{\circ}41'26.8''$. The spectrum covered C, X, Ku, K, and Ka bands, spanning 29~GHz in bandwidth with a resolution of 1.4~kHz. The cyanotriphenylene spectra were first simulated at the typical TMC-1 conditions ($T = 7$~K, $v_{LSR} = 5.8~\mathrm{km~s^{-1}}$, $\Delta v = 0.4~\mathrm{km~s^{-1}}$) with the \texttt{molsim} package \citep{molsim}. Under these conditions, the spectral envelope follows the Boltzmann distribution, gathering the strongest spectral lines of cyanotriphenylenes at X-band, where GOTHAM data have the lowest noise level ($\sim 2$~mK). However, none of the individual spectral lines were distinguishable under the current noise level. The $3\sigma$ upper limit of column density was determined to be $2.19\times10^{12}~\text{cm}^{-2}$ for 1-CNT (based on the line with the highest SNR located at 10265.891~MHz) and $1.54\times10^{12}~\text{cm}^{-2}$ for 2-CNT (based on the line with the highest SNR located at 10201.196~MHz).

To obtain a more constraining limit than that from individual transitions, we performed velocity stacking and matched-filter analysis using the 150 transitions predicted to have the highest SNR, following \citealt{stacking}. Because the systemic velocity of TMC-1 is known, the search was evaluated at the expected velocity channel rather than by scanning for arbitrary peaks. No statistically significant matched-filter response higher than $3 \sigma$ was found for either cyanotriphenylene isomer.

We estimated the sensitivity of this analysis through injection-recovery simulations by injecting molecular signals of different column densities into synthetic spectra reproducing the local noise properties of the GOTHAM data. Column densities of $3.0\times10^{11}~\text{cm}^{-2}$ for 1-CNT and $2.5\times10^{11}~\text{cm}^{-2}$ for 2-CNT produced $>3\sigma$ matched-filter responses in 98.91\% and 99.75\% of the trials, respectively. Details are provided in the Supporting Information.

Cyanotriphenylene has only two isomers, and these have larger or comparable permanent electric dipole moments when compared to three cyanopyrene isomers, which were detected with strong signals (including individual observable single-line transitions) with $1.52\times10^{12}~\text{cm}^{-2}$, $0.84\times10^{12}~\text{cm}^{-2}$, and $1.33\times10^{12}~\text{cm}^{-2}$ for column densities of 1-, 2, and 4-cyanopyrene, respectively \citep{cyanopyrene,cyanopyrene2}. Therefore, it is reasonable to conclude that cyanotriphenylenes are missing or at least have a significantly lower abundance than cyanopyrenes in TMC-1. From this and the barrierless formation route to CNTs, we suggest that triphenylene may be relatively suppressed compared with the
pyrene and coronene frameworks traced by their cyano derivatives in TMC-1, provided that CN tagging proceeds with broadly comparable efficiencies among these PAHs.

\begin{figure*}[t!]
\begin{center}
\includegraphics[width=\textwidth]{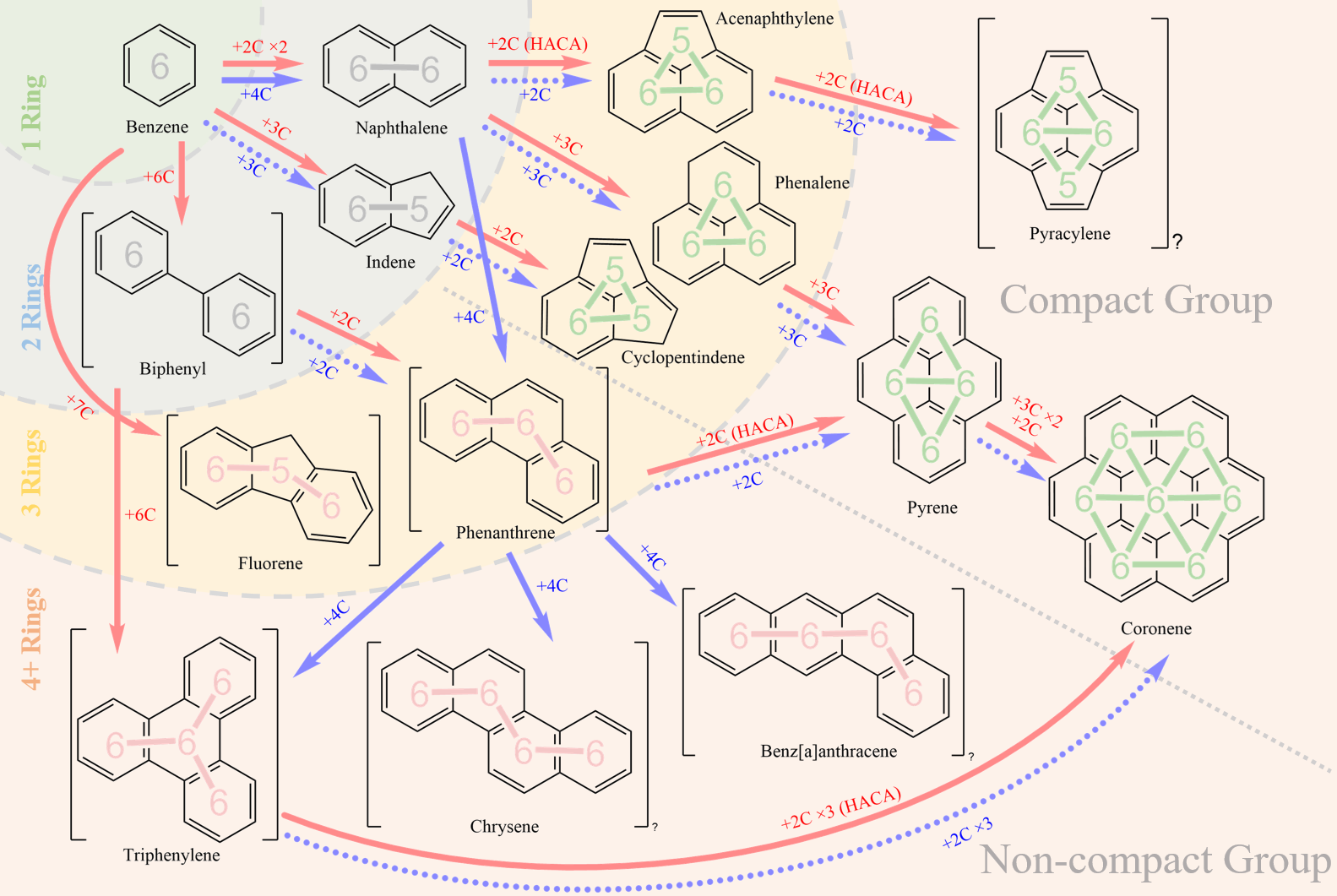}    
\caption{Background colors indicate ring number. Species enclosed in square brackets, or their CN-substituted analogues, have not been detected in TMC-1; question marks indicate structures whose CN derivatives lack laboratory rotational spectra. Blue arrows denote barrierless known HAVA-type +4C growth pathways viable under cold interstellar conditions; blue dashed arrows indicate other potential barrierless routes from the neutral PAH reacting with small radical species. Red arrows indicate reactions expected to involve activation barriers and thus to require warmer environments. Ring-node diagrams (dualist graph) classify each PAH by treating aromatic rings as nodes labeled by carbon count and connecting rings that share an edge. PAHs whose ring-node graphs contain closed loops are classified as compact and shown in green, whereas non-compact PAHs are shown in red. Species with two or fewer rings cannot form closed loops in this representation and are shown in gray. The gray dotted line separates the compact and non-compact groups.}
\label{fig:structure}
\end{center}
\end{figure*}

\subsection{Mechanistic Implications of the Missing Cyanotriphenylene}\label{subsec:discussion}
The following discussion is organized into three parts. We first assess whether cyanotriphenylenes can serve as tracers of triphenylene using quantum-chemical calculations and discuss the resulting ``missing triphenylene'' interpretation. We then use a topological graph representation to place the non-detection within the broader PAH inventory of TMC-1. Finally, we consider two kinetic scenarios that may contribute to the apparent preference for compact structures: sequential growth decay and fast consumption of non-compact PAHs.

\subsubsection{Tracer Validity and Abundance Inference}
Before further discussion of the non-detection, the feasibility of choosing cyanotriphenylenes as tracers of triphenylene was first examined computationally. In Fig.~\ref{fig:PES}, high-level ab initio calculations characterize the PES of two pathways that generate 1-CNT or 2-CNT when triphenylene reacts with the CN radical. Both pathways involve a barrierless first step and a submerged transition state for the second step, indicating that this gas-phase reaction is kinetically accessible at typical interstellar medium (ISM) temperatures \citep{gasphasekinetics}. We also calculated the PES for the generation of isocyanotriphenylene (1-iCNT and 2-iCNT), and both pathways show positive activation barriers for the transition states ($>$35 kJ mol$^{-1}$ for TS2), making them unfavorable under cold ISM conditions. Together with prior measurements of \ce{CN + benzene} \citep{CN+bezene} and calculations on \ce{CN + pyrene} \citep{cyanopyrene2}, these PES results support the use of cyanotriphenylenes as viable tracers of triphenylene: CN addition to triphenylene is 
expected to be kinetically accessible under cold-cloud conditions.

The non-detection of the cyanotriphenylene isomers suggests, rather than directly demonstrates, that triphenylene may be relatively scarce in TMC-1 compared with other large PAHs such as pyrene, provided that CN tagging proceeds with broadly comparable efficiencies among these PAHs. This possible relative underabundance motivates what we refer to here as the ``missing triphenylene'' puzzle. From a thermodynamic perspective, triphenylene would be expected to be a favorable product because it is the most stable member of the four-ring unsubstituted benzenoid \ce{C18H12} isomer family, as shown in Fig.~\ref{fig:energies}. Although the full \ce{C18H12} isomer set has not yet been surveyed astronomically, the pyrene carbon skeleton has already been identified indirectly in TMC-1 through detections of 1-, 2-, and 4-cyanopyrene \citep{cyanopyrene,cyanopyrene2}. Pyrene and triphenylene are not isomers, so their stabilities cannot be compared directly through relative energies alone; however, triphenylene has both a higher first ionization potential and a lower electron affinity \citep{electronicstability}, indicating greater electronic stability that may favor survival under interstellar irradiation. Taken together, these observations suggest that the PAH inventory in TMC-1 cannot be rationalized solely on the basis of thermodynamic stability and is consistent with important roles for formation and destruction kinetics. Similar departures from thermodynamic control are already well established for smaller interstellar molecules, including the \ce{C3H2O} isomers in Sgr B2(N) \citep{H2C3O1} and the absence of the most stable \ce{C3H6O2} isomer, propionic acid, in G+0.693-0.027 \citep{C3H6O2}. The present results suggest that analogous considerations may also be relevant to large PAHs in TMC-1.

\subsubsection{Topological Graph and Structure Selectivity}

From the perspective of structural features, the absence of triphenylene points most directly to an unfavorable role for bay-containing local motifs in TMC-1. Triphenylene was sometimes grouped with compact PAHs in the literature \citep{Langhoff1996}, so its non-detection might at first seem to argue against any compactness preference in this source. However, this conclusion would be premature because compactness itself lacks a unique and transferable definition across different PAH skeletons. For example, other studies regard triphenylene as a less compact PAH \citep{triphenylenelesscompact}. To address this ambiguity, we adapted an old topological classification, dualist graph \citep{Anatomy}, as introduced by Fig.~\ref{fig:structure}. The dualist graph was originally developed to categorize benzenoid hydrocarbons and to analyze their molecular formulas and thermodynamic properties. Here, we extend this representation to PAHs containing non-benzenoid rings for astrochemical applications. Specifically, it is a ring-node representation in which each fused ring is treated as a node and compactness is defined through the topology of the resulting ring graph. Similar classification was named as catacondensed/pericondensed \citep{Enumeration} that corresponding to non-compact/compact here. Within this classification, triphenylene falls into the non-compact cluster: it contains three branching bay regions and does not form the closed-loop ring-node topology that characterizes the compact group. In this sense, non-compactness and the presence of bay regions are not separate features, but rather two manifestations of the same topological clustering. Under this framework, the PAHs detected in TMC-1 are predominantly compact, whereas the missing species, including fluorene, phenanthrene, and triphenylene, cluster on the non-compact side. This classification also suggests clear predictions for other PAHs that remain untested experimentally: pyracylene should be a promising compact target, whereas the non-compact isomers chrysene and benz[a]anthracene are expected to be much less abundant. As a result, these species become particularly informative targets for testing the proposed topology-dependent selectivity. Future laboratory rotational characterization of suitable polar derivatives, followed by sensitive astronomical searches toward TMC-1, could determine whether the current compact/non-compact tendency persists as the interstellar PAH inventory expands.

The same classification is consistent with the broader chemical character of TMC-1, where highly unsaturated carbon species, such as polyynes, are abundant. Compact PAH growth can be viewed as a carbon-efficient mode of aromatization, because forming a new ring across two pre-existing fused rings closes a compact loop in the ring-node graph, whereas peripheral ring extension simply elongates the skeleton and forms non-compact PAHs. The former process increases ring number with fewer additional peripheral C–H sites, leading to a higher effective C/H ratio and a more extensively conjugated carbon framework. Thus, compactness, as defined here, is not merely a geometric descriptor but also a chemical signature of dehydrogenated, carbon-rich growth. 

\subsubsection{Formation-Side Scenario: Sequential Growth Decay}

To understand the kinetic origin of this topological clustering, we first consider the problem from the formation side by examining plausible bottom-up growth pathways among the PAHs summarized in Fig.~\ref{fig:structure}. The relevant ring-growth mechanisms include Hydrogen Abstraction-\ce{C2H2} Addition (HACA, 2-carbon addition, \citealt{HACA1}), Hydrogen Abstraction-\ce{C3H4} Addition (3-carbon addition, \citealt{C3addition1}), Hydrogen Abstraction-Vinylacetylene Addition (HAVA, 4-carbon addition, \citealt{HAVA1}), and Phenyl Addition-Dehydro Cyclization (PAC, 6-carbon addition, \citealt{PAC}). Of particular relevance here is that many HAVA pathways are barrierless according to quantum chemical calculations and laboratory studies, making it a potentially viable route to PAH growth under cold-cloud conditions \citep{review}. However, within the set of pathways relevant to the present discussion, HAVA growth from smaller detected PAHs tends to access bay-containing, non-compact products that are missing or underabundant in TMC-1, including phenanthrene from naphthalene \citep{nath2phen} and triphenylene, chrysene, and benz[a]anthracene from phenanthrene \citep{gasphase,phen2chry}. By contrast, several 2- and 3-carbon ring-growth pathways can generate compact closed-loop topologies in the ring-node sense, because the newly formed five- or six-membered ring shares edges with existing rings and closes the ring graph. Examples include 2-carbon growth from naphthalene to acenaphthylene \citep{Nath2Acey} and 3-carbon growth from naphthalene to phenalene \citep{Nath2Phan}. Repeated 2-carbon addition to naphthalene can further generate pyracylene, whereas repeated 3-carbon growth followed by hydrogen loss can produce pyrene. Although the HACA mechanism appears chemically plausible as a route to several detected compact PAHs, both calculations and laboratory studies generally indicate positive activation barriers for these reactions. Such barriers make HACA-driven growth difficult to reconcile with the $\sim$7 K kinetic temperature of TMC-1. By contrast, the bay-containing PAHs most readily accessible through cold-favorable HAVA pathways remain missing. This tension highlights a major mismatch between the formally available reaction network and the PAH inventory actually observed in the source.

To account for the absence of these HAVA-accessible products including triphenylene, we propose a ``sequential growth decay'' hypothesis. In this picture, a multistage HAVA ring-growth chain can in principle proceed through several successive steps, analogous to the extension of cyanopolyyne chains from \ce{HC3N} to \ce{HC11N} \citep{stacking}, but the abundance of the products decreases rapidly at each stage. Such a formation bottleneck could explain why triphenylene and other non-compact PAHs such as phenanthrene remain below the current detection limit. Notably, the route from indene to fluorene is unfeasible for the HAVA mechanism \citep{indeneHAVA}, so the non-detection of the fluorene framework is not included in this specific argument.

\subsubsection{Destruction-Side Scenario: Fast Consumption}

While the ``sequential growth decay hypothesis'' attributes the non-detections to formation bottlenecks, an alternative scenario called ``the fast-consumption hypothesis'' suggests they are actively depleted by subsequent reactions. Although extensively studied ring-growth mechanisms typically involve hydrogen abstraction as the initial step, which is likely hindered in cold, dark molecular clouds because of the low thermal energy available, TMC-1 may still contain a sufficient abundance of unsaturated small radicals capable of participating in low-temperature ring-growth chemistry, such as \ce{CH} \citep{CH}, \ce{C2} \citep{C2}, \ce{C2H} \citep{C2HtoBenzene}, and \ce{C3} \citep{C3}. Among them, reactions involving \ce{C2H} (the ethynyl radical), also referred to as the ethynyl addition mechanism (EAM), have been proposed and calculated as a potential low-temperature substitute for HACA \citep{ethynyladdition}. Additionally, the detection of ethynyl-substituted molecules in TMC-1, such as ethynylcyclopentadiene \citep{ethynylcyclopentadiene} and ethynylbenzene \citep{ethynylbenzene}, indirectly supports a relatively high abundance of ethynyl radicals in this source, consistent with the observed abundance ratio CCH/CN $\approx 10$ in the cyanopolyyne peak \citep{CCHCNratio,ethynylcyclopentadiene}. In particular, if ring-growth channels following \ce{C2H} addition at bay regions are competitive, such reactions could contribute to the consumption of non-compact PAHs and their conversion into compact PAHs (e.g., from phenanthrene to pyrene, or from triphenylene to coronene). However, the product branching between H-loss, rearrangement, and ring-closure channels under cold-cloud conditions has not yet been established in published work. This fast-consumption hypothesis offers an explanation for the low abundance of non-compact species while accounting for the high column densities measured for cyano derivatives of compact PAH frameworks such as pyrene and coronene \citep{cyanopyrene,cyanopyrene2,cyanocoronene}. Extending this picture to three-carbon radicals further broadens the set of plausible low-temperature growth channels: \ce{c-C3H} \citep{l-C3H}, \ce{l-C3H} \citep{l-C3H}, \ce{c-C3H2}, \ce{l-C3H2} \citep{C3H2,C3H2-2}, and even \ce{C3H3} (the propargyl radical, also denoted \ce{CH2CCH} \citep{C3H3}) have all been detected in TMC-1 and may therefore motivate additional three-carbon ring-growth pathways. However, resonantly stabilized radicals such as propargyl cannot be assumed to undergo barrierless addition to closed-shell PAHs. DFT calculations of propargyl addition to naphthalene found pathway-dependent energetics, including positive barriers for relevant addition and rearrangement channels, \cite{Raj2014} suggesting that resonance stabilization may restrict some low-temperature ring-growth pathways, although the quantitative barrier heights require validation with higher-level theoretical methods. Accordingly, the fast-consumption scenario is retained here as a testable hypothesis rather than an established mechanism.

More importantly, this scenario does not rely on closed-shell two- or three-carbon precursors as the key ring-growth reagents, unlike the classic HACA mechanism. At a higher mechanistic level, the key distinction is whether H loss occurs first from a small precursor or from a large aromatic molecule: EAM-type pathways lose hydrogen from the smaller reactant to form reactive radicals, whereas HACA-type pathways require prior hydrogen abstraction from the larger PAH. Experimental and modeling studies of PAH dehydrogenation further show that once aliphatic-like C-H sites are formed, H loss proceeds more readily than from purely aromatic edge sites \citep{Hloss}, which further favors radical generation on the small-reactant side. Because the parent small molecules are generally more abundant than large aromatic species in TMC-1 and are also more easily converted into radicals by H loss, the corresponding reactive radical pool is expected to be larger, making EAM-type pathways kinetically more favorable. This scenario therefore offers a way to rationalize not only the disappearance of bay-containing, non-compact intermediates, but also the formation of detected compact PAHs such as acenaphthylene under conditions where conventional HACA-type chemistry is unlikely to operate efficiently at very low temperature. Consequently, further computational and experimental studies on the roles of these unsaturated small radicals are necessary.

\section{Conclusions}
In this work, we synthesized both 1-cyanotriphenylene and 2-cyanotriphenylene, and measured their microwave rotational spectra as polar tracers for interstellar triphenylene. An astronomical search in TMC-1 based on these spectroscopic parameters yielded no statistically significant detection of the cyanotriphenylenes, motivating a chemically informative ``missing triphenylene'' hypothesis for a thermodynamically favored PAH framework. This inference assumes that CN substitution occurs with broadly similar efficiencies among PAHs. To interpret this result, we introduced a ring-node representation that defines compactness topologically and places triphenylene in the non-compact, bay-containing cluster despite its frequent treatment as a compact PAH in the conventional literature. Within this framework, most PAHs detected in TMC-1 are compact, whereas the undetected candidates considered here are non-compact. This contrast suggests that topology may help shape the observed PAH inventory, which cannot be explained by relative thermodynamic stability alone. We further suggest two chemical origins for this selectivity: a sequential growth decay along HAVA-accessible formation pathways and a fast-consumption process in which reactive unsaturated radicals could efficiently transform bay regions into more compact structures. Future laboratory and computational studies on additional non-compact PAHs and on radical reactions at bay-region sites will be essential for distinguishing between these mechanisms.

\section{Data Availability} \label{sec:data}
The original spectroscopic files of 1-CNT and 2-CNT, including lin files from experimental measurement, var/par files for fitted spectroscopic parameters, int files for simulation parameters, qpart files for the partition functions, piform files for the reformatted fitting results, and cat files for observational searching, are all available on Zenodo under an open-source Creative Commons Attribution license: 
\dataset[doi:10.5281/zenodo.22967120]{https://doi.org/10.5281/zenodo.22967120}.
\\
\section*{AI Use Disclosure}

OpenAI's ChatGPT was used under author supervision to assist with English-language proofreading and stylistic editing of author-written text, to improve clarity and readability. It was not used to generate research data or determine scientific conclusions. All suggested changes were reviewed and approved by the authors, who take full responsibility for the final manuscript.

\begin{acknowledgments}
    We thank Siyuan Gong for guidance and assistance with the nitrile synthesis. We gratefully acknowledge support from NSF grants AST-1908576, AST-2205126, and AST-2307137 and Schmidt Family Futures. G.W. and B.A.M. acknowledge the support of the Arnold and Mabel Beckman Foundation Beckman Young Investigator Award. The National Radio Astronomy Observatory is a facility of the National Science Foundation operated under cooperative agreement by Associated Universities, Inc. The Green Bank Observatory is a facility of the National Science Foundation operated under cooperative agreement by Associated Universities, Inc.  T.H.S. acknowledges the support of the Canadian Space Agency (CSA) through grant 24AO3UBC14.
    
\end{acknowledgments}
\facility{GBT - Green Bank Telescope.}

\appendix
\section{The Synthesis of Cyanotriphenylenes}
\label{app:synthesis}

\begin{figure}[htb!]
    \centering
    \includegraphics[width=0.4\linewidth]{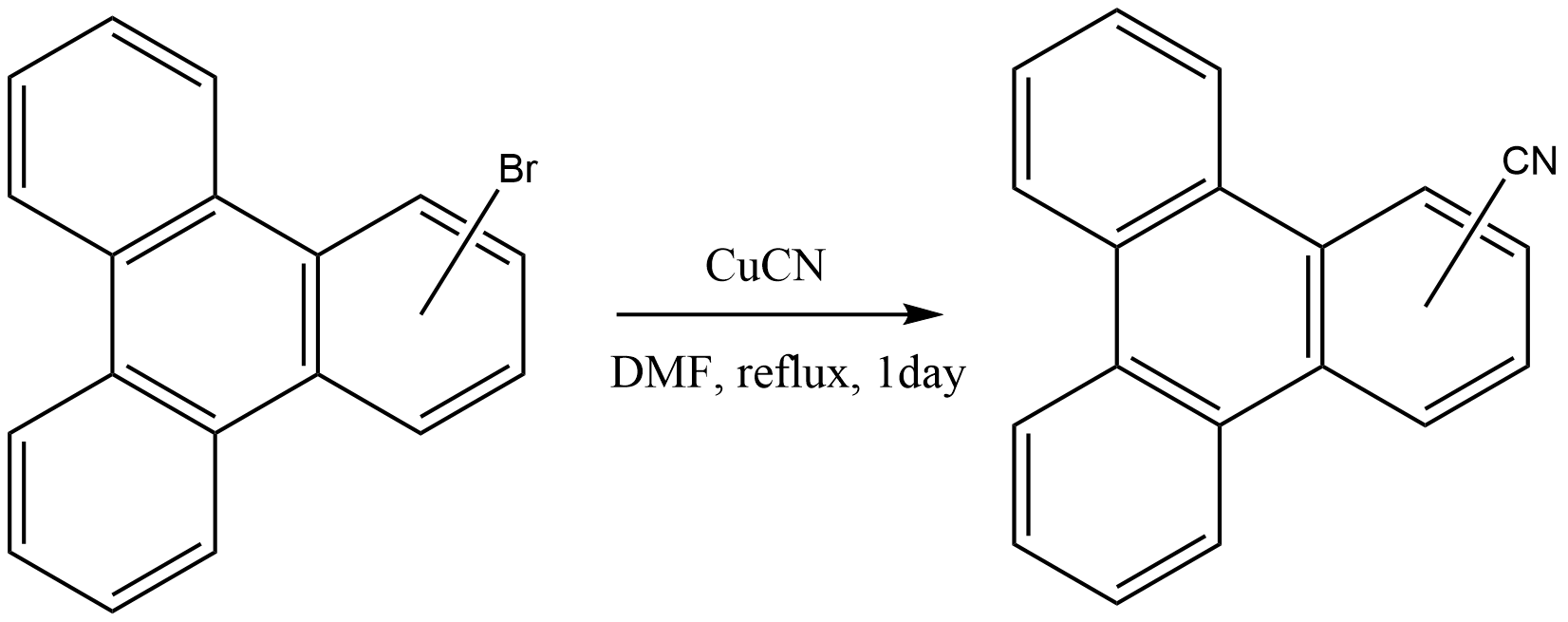}
    \caption{Full synthesis route from bromotriphenylenes to cyanotriphenylenes.}
    \label{fig:synthesis}
\end{figure}

The full synthesis route from bromotriphenylenes to cyanotriphenylenes is depicted in Fig.~\ref{fig:synthesis} and described in the following.

The Rosenmund–von Braun reaction conditions were adapted from a reported procedure for 4-cyanopyrene synthesis \citep{cyanopyrene2}. Reagents, including 1-bromotriphenylene, 2-bromotriphenylene (A2B Chem), cuprous cyanide (CuCN, Sigma-Aldrich), and $N,N$-dimethylformamide (DMF, Aldrich) were purchased and used as received. DMF was dried over $3\,\text{\AA}$ molecular sieves overnight prior to use. 

\subsection{The Synthesis of 1-cyanotriphenylene}
To a dried 100 mL three-neck round-bottom flask equipped with a reflux condenser were added CuCN ($0.27\,\text{g}$, $3\,\text{mmol}$) and 1-bromotriphenylene ($0.46\,\text{g}$, $1.5\,\text{mmol}$). The system was evacuated and backfilled with nitrogen three times. $40\,\text{mL}$ anhydrous DMF was added via syringe under the protection of nitrogen, and the mixture was heated to reflux for $24\,\text{h}$. After cooling to room temperature, the reaction was quenched by the addition of $100\,\text{mL}$ deionized water. The aqueous layer was extracted with ethyl acetate 3 times. The combined organic layers were washed with brine, dried over anhydrous $\text{Na}_2\text{SO}_4$, filtered, and concentrated under reduced pressure. The crude residue was purified by flash column chromatography on silica gel (petroleum ether/ethyl acetate = 20:1) to produce 1-cyanotriphenylene as light yellow powder exhibiting dark blue fluorescence under a UV lamp. 

\begin{figure}
    \centering
    \includegraphics[width=\linewidth]{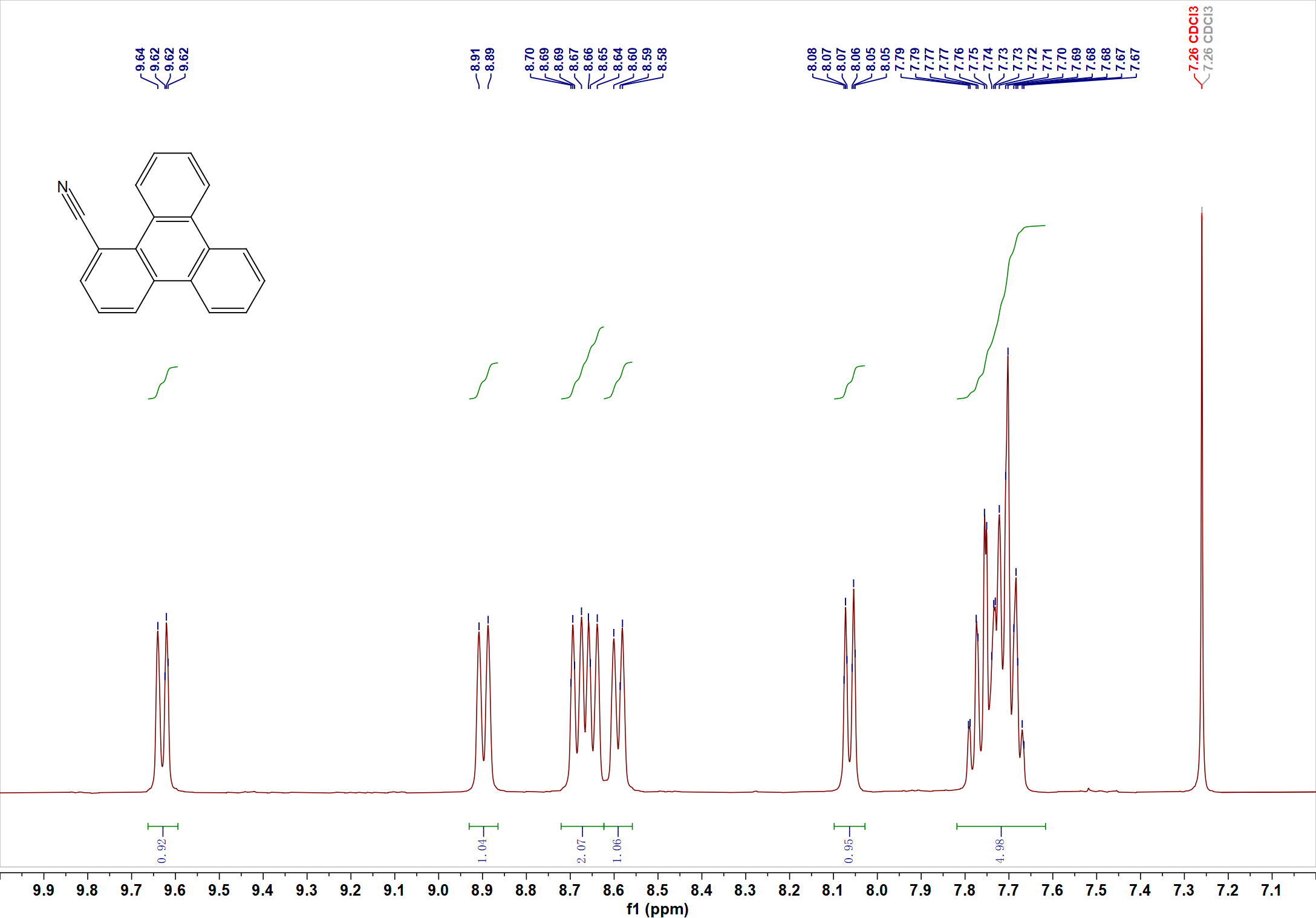}
    \caption{$^1$H NMR spectrum of 1-cyanotriphenylene.}
    \label{fig:1HNMR-1-CNT}
\end{figure}

\noindent \textbf{$^{1}$H NMR} (400 MHz, CDCl$_{3}$) $\delta$ 9.70–-9.58 (m, 1H), 8.90 (d, $J = 8.3$\,Hz, 1H), 8.72--8.62 (m, 2H), 8.59 (d, $J = 7.9$\,Hz, 1H), 8.06 (dd, $J = 7.4, 1.3$\,Hz, 1H), 7.84--7.63 (m, 5H) shown in Figure.~\ref{fig:1HNMR-1-CNT}. 

\begin{figure}
    \centering
    \includegraphics[width=\linewidth]{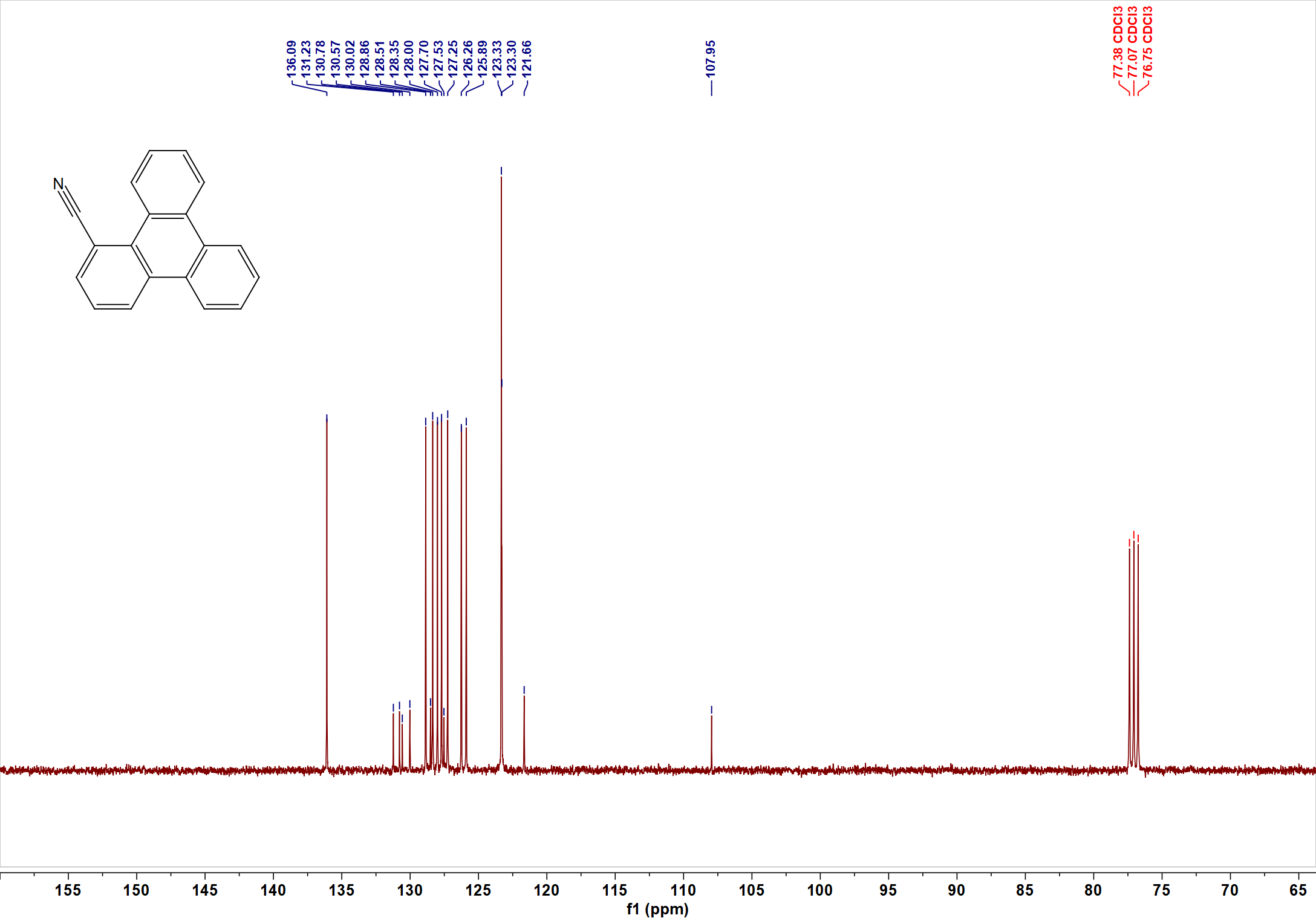}
    \caption{$^{13}$C NMR spectrum of 1-cyanotriphenylene.}
    \label{fig:13CNMR-1-CNT}
\end{figure}

\noindent \textbf{$^{13}$C NMR} (101 MHz, CDCl$_{3}$) $\delta$ 136.09, 131.23, 130.78, 130.57, 130.02, 128.86, 128.51, 128.35, 128.00, 127.70, 127.53, 127.25, 126.26, 125.89, 123.33, 123.30, 121.66, 107.95, as shown in Figure.~\ref{fig:13CNMR-1-CNT}.  These data are consistent with those reported in the literature \cite{1-CNTNMR}.

\subsection{The Synthesis of 2-cyanotriphenylene}
Used similar methods as described for the synthesis of 1-CNT, but added different amount of CuCN ($2.70\,\text{g}$, $30\,\text{mmol}$), 2-bromotriphenylene ($2.76\,\text{g}$, $9\,\text{mmol}$), and anhydrous DMF ($30\,\text{mL}$) for the scaled test. The crude residue was purified by flash column chromatography on silica gel (petroleum ether/ethyl acetate = 20:1) to produce 2-cyanotriphenylene as needle-shaped crystals exhibiting blue fluorescence under a UV lamp. 

\begin{figure}
    \centering
    \includegraphics[width=\linewidth]{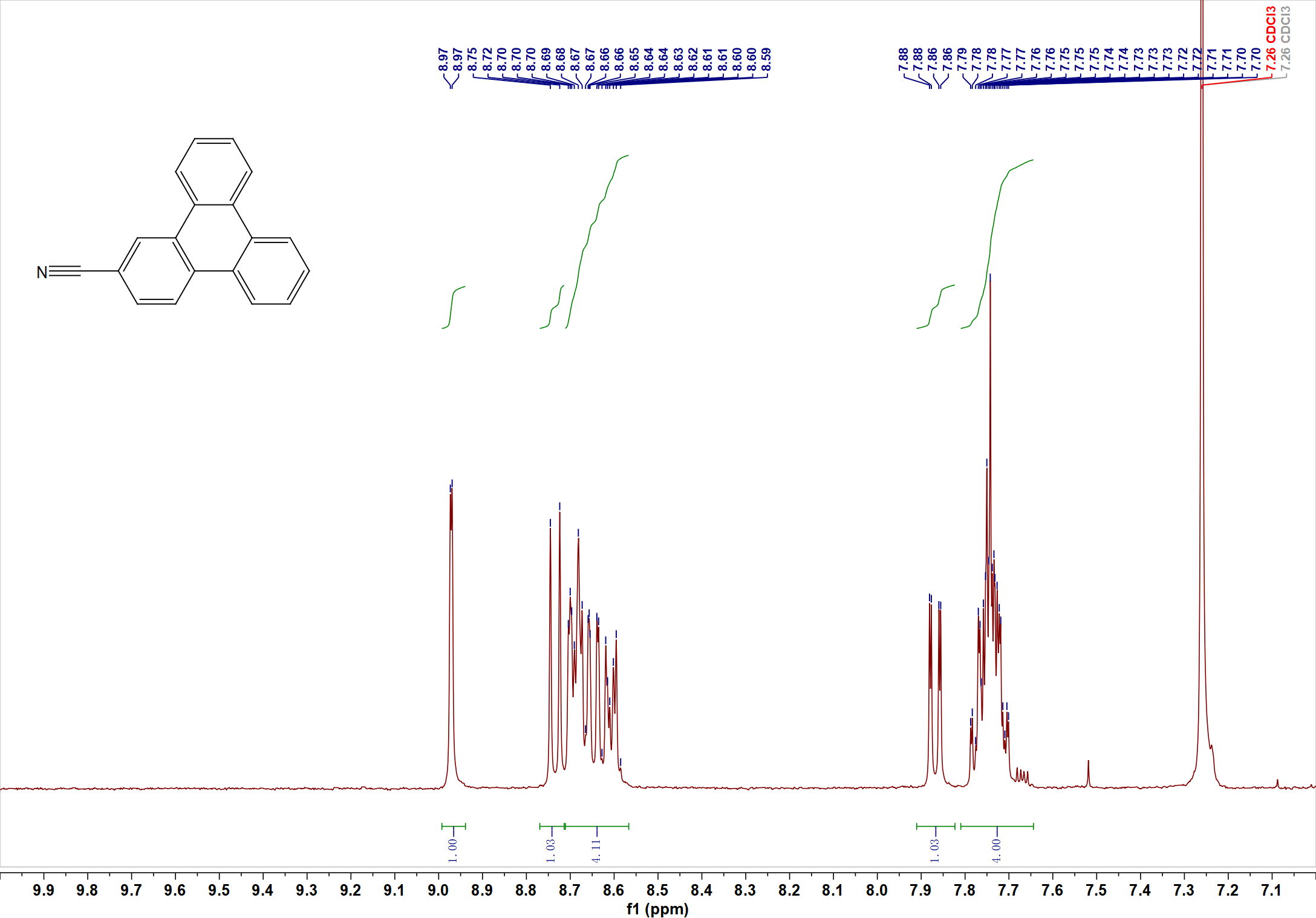}
    \caption{$^1$H NMR spectrum of 2-cyanotriphenylene.}
    \label{fig:1HNMR-2-CNT}
\end{figure}

\noindent \textbf{$^{1}$H NMR} (400 MHz, CDCl$_{3}$) $\delta$ 8.97 (d, $J = 1.7$\,Hz, 1H), 8.73 (d, $J = 8.5$\,Hz, 1H), 8.71--8.50 (m, 4H), 7.87 (dd, $J = 8.5, 1.7$\,Hz, 1H), 7.80--7.69 (m, 4H), shown in Figure.~\ref{fig:1HNMR-2-CNT}. 

\begin{figure}
    \centering
    \includegraphics[width=\linewidth]{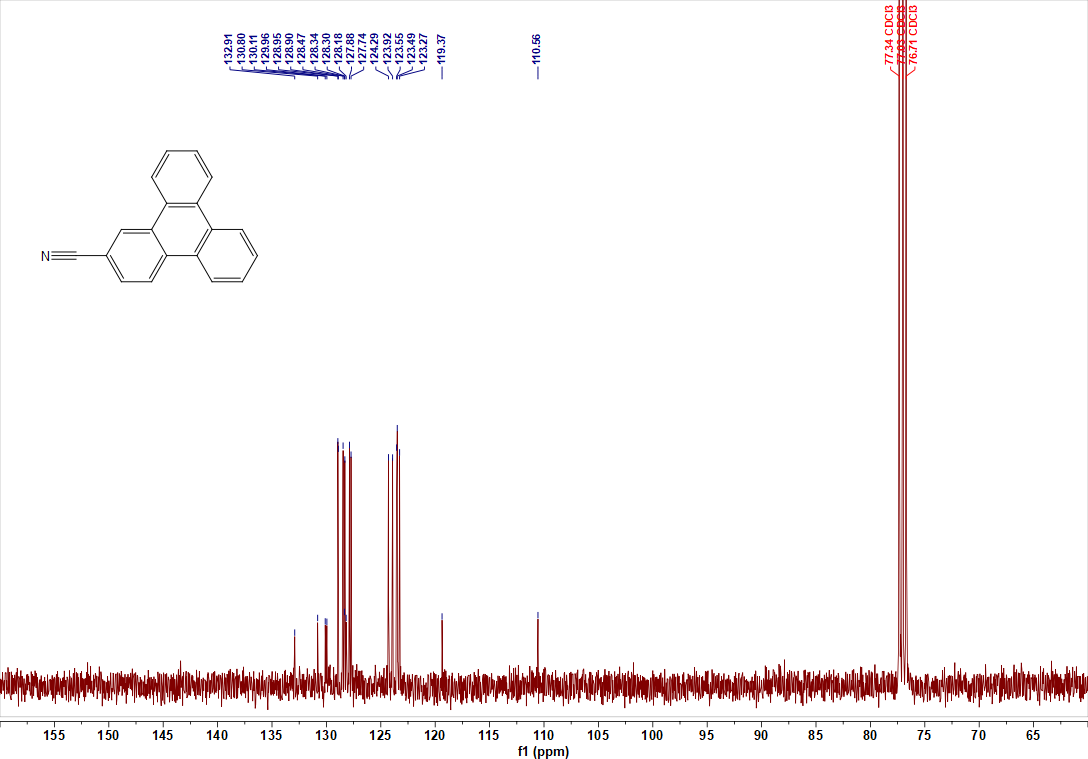}
    \caption{$^{13}$C NMR spectrum of 2-cyanotriphenylene.}
    \label{fig:13CNMR-2-CNT}
\end{figure}

\noindent \textbf{$^{13}$C NMR} (101 MHz, CDCl$_{3}$) $\delta$ 132.91, 130.80, 130.11, 129.96, 128.95, 128.90, 128.47, 128.34, 128.30, 128.18, 127.88, 127.74, 124.29, 123.92, 123.55, 123.49, 123.27, 119.37, 110.56, as shown in Figure.~\ref{fig:13CNMR-2-CNT}. These data are consistent with those reported in the literature \cite{NMR1,NMR2}.

\clearpage

\section{The statistical analysis of upper limits for cyanotriphenylenes}

To establish a more constrained and observation-specific upper limit estimate for the two cyanotriphenylene isomers, we performed a noise-injection and recovery analysis using synthetic spectra that reproduce the frequency-dependent noise properties of the GOTHAM DR5.3 observations. The molecular emission was simulated using the four-component TMC-1 source model adopted for benzonitrile\citep{DR5.3}, with the velocity components, source sizes, excitation temperature, and linewidths listed in Table~\ref{table:sourceparams}. For each trial column density, the total column density was divided equally among the four velocity components. This treatment preserves the multi-component velocity structure of TMC-1 while allowing the total column density of each cyanotriphenylene isomer to be varied as the only free scaling parameter.

\begin{figure}
    \centering
    \includegraphics[width=\linewidth]{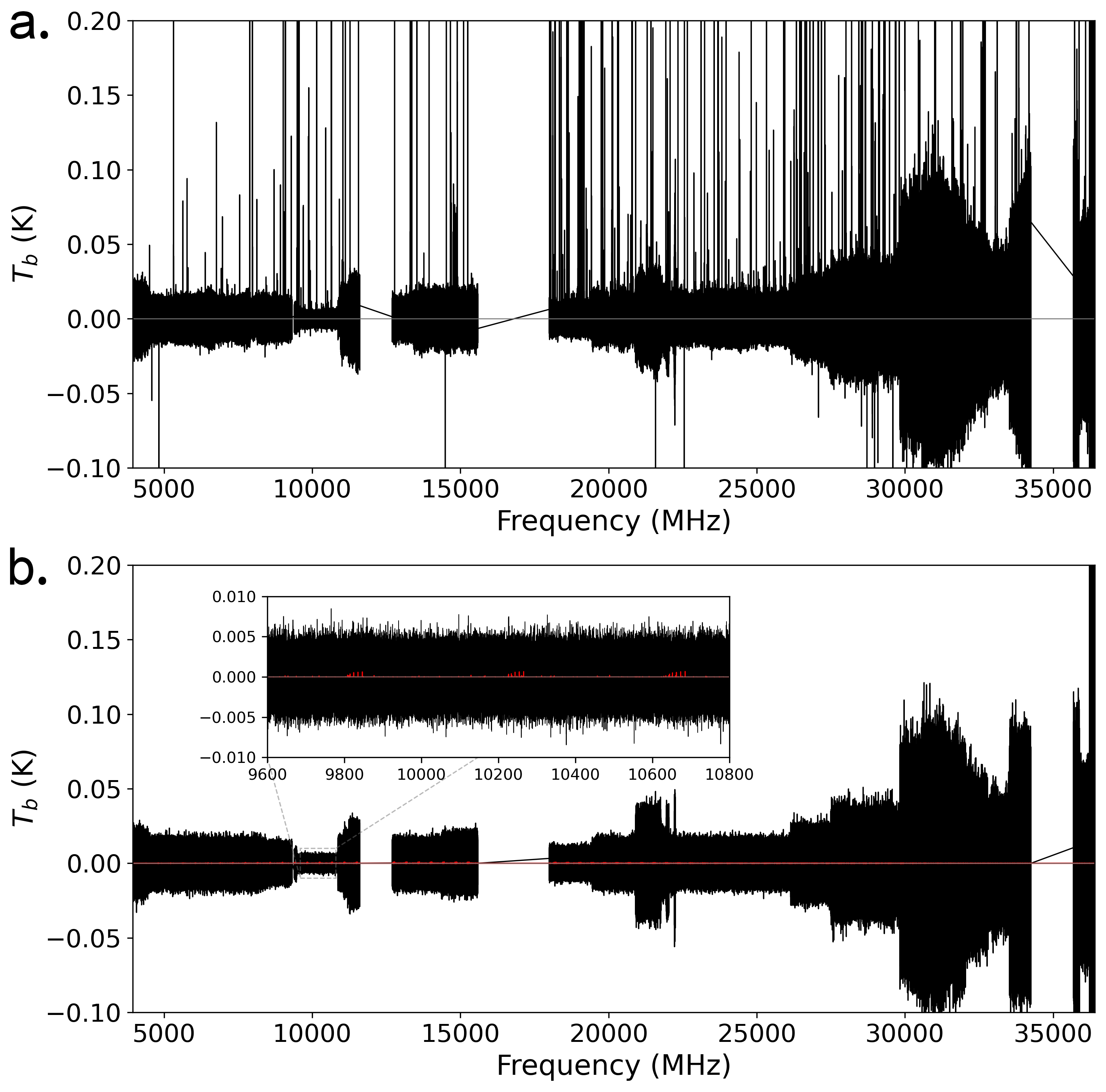}
    \caption{Noise-injection procedure used for the stacking upper-limit analysis. (a) Full DR5.3 GOTHAM spectrum used as the observational reference. The frequency-dependent noise structure and flagged spectral gaps are preserved across the survey bandwidth. (b) Synthetic noise realization generated over the same frequency grid using the local RMS values measured from the DR5.3 spectrum within the predefined frequency intervals. The simulated four-component 1-CNT spectrum with $3.5\times10^{11}~\text{cm}^{-2}$ total column density is overlaid in red to illustrate the injected signal used in the recovery tests. The inset shows an expanded view of the 9600-10800 MHz region, highlighting that the injected molecular features are weak relative to the local noise.}
    \label{fig:noisesimulation}
\end{figure}

The synthetic noise spectra were generated on the same frequency grid as the DR5.3 GOTHAM data. In order to retain the observational characteristics relevant for stacking, the local Root-Mean-Square (RMS) measured from the observed spectrum was used to set the amplitude of the synthetic Gaussian noise within predefined frequency intervals, while the flagged spectral gaps and frequency-dependent sensitivity variations of the survey were preserved. This procedure is shown in Figure~\ref{fig:noisesimulation}: the full DR5.3 spectrum provides the observational noise reference, and the corresponding synthetic realization reproduces the same broadband noise structure and gap regions. Simulated molecular spectra were then injected into these noise realizations at a series of trial column densities. The inset in Figure~\ref{fig:noisesimulation} shows that the injected cyanotriphenylene features are weak compared with the local noise in individual spectral channels, motivating the use of stacking and matched filtering rather than single-transition searches.

\begin{figure}
    \centering
    \includegraphics[width=\linewidth]{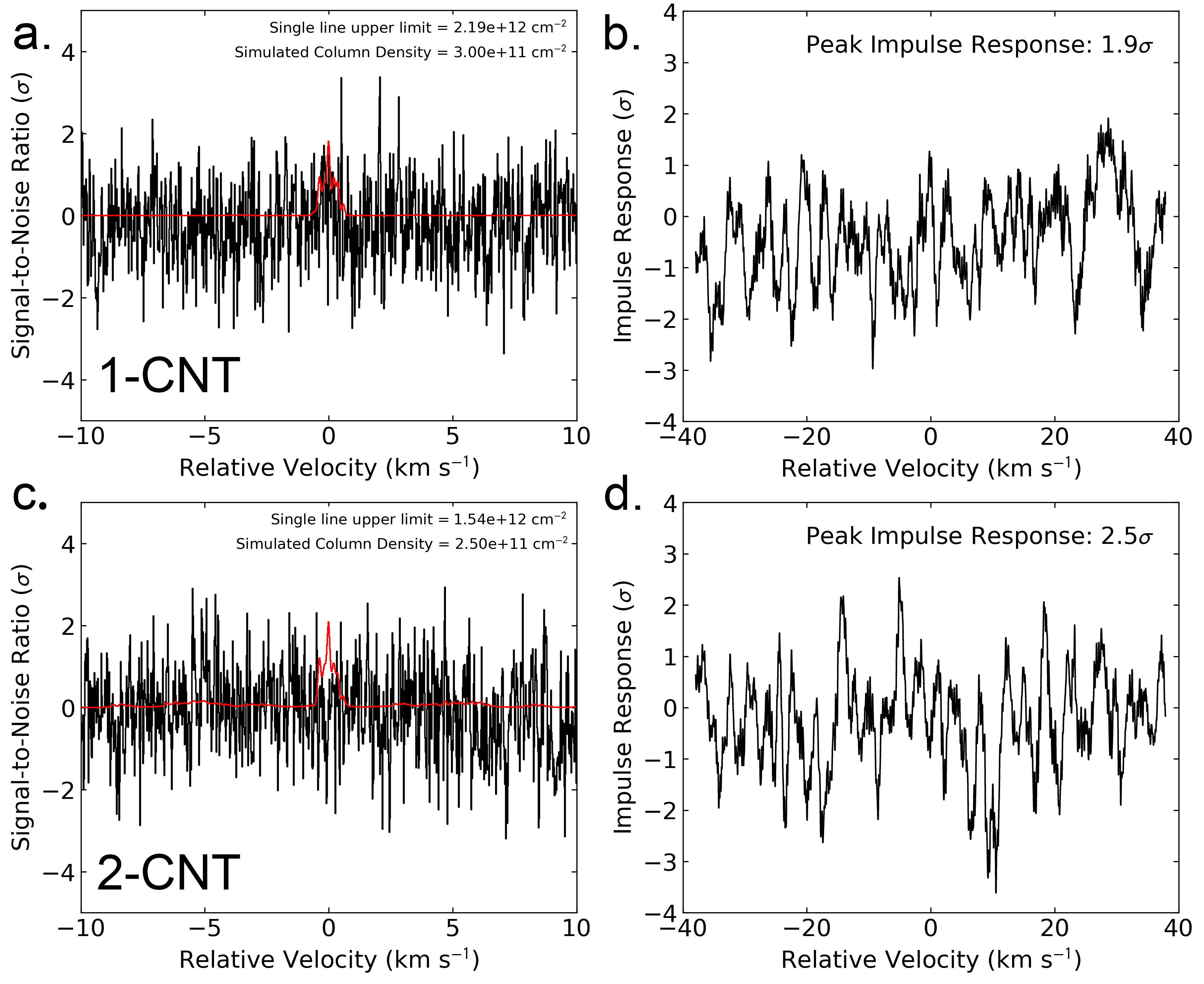}
    \caption{The spectral line stacking and matched filtering result of cyanotriphenylenes towards TMC-1 with GOTHAM DR5.3 observational data. The intensity of the red lines in figures (a) and (c) corresponds to the stacking result of 150 highest SNR 4-component molecular peaks with column density simulated at the stacking upper limit. No peaks located at around 0 $\mathrm{km\,s^{-1}}$ are higher than $2 \sigma$ in the impulse response figures (b) and (d).}
    \label{fig:nondetectioin}
\end{figure}

For each injected spectrum, we applied the same spectral line stacking and matched-filtering procedure used to assess the observational data. The stacking template was constructed from the 150 transitions with the highest predicted signal-to-noise ratios for each isomer, including the four velocity components in the source model. The stacked spectrum was then cross-correlated with the corresponding simulated molecular template, and the peak impulse response was recorded in units of $\sigma$. Further details of the stacking and matched-filtering formalism are described in \citet{stacking}. As a reference, the stacking and matched-filtering results for the actual DR5.3 observations are shown in Figure~\ref{fig:nondetectioin}. No statistically significant feature is present at the expected systemic velocity in the stacked spectra of either isomer, and the corresponding matched-filter impulse responses do not show a peak at $\sim0~\mathrm{km~s^{-1}}$ exceeding $2\sigma$. The single-line upper limits are therefore substantially less constraining than the limits obtained from the combined stacking analysis.

\begin{figure}
    \centering
    \includegraphics[width=\linewidth]{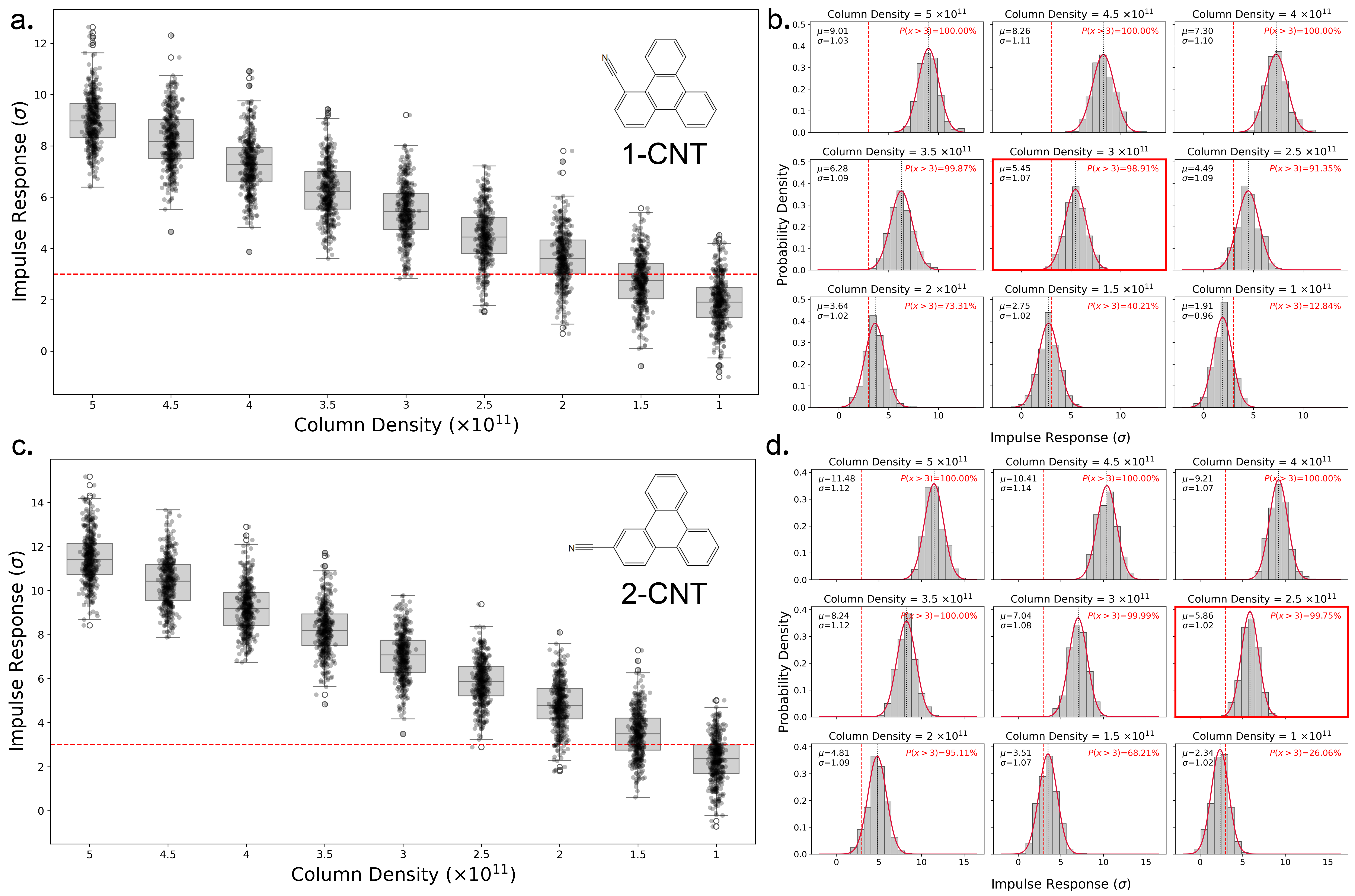}
    \caption{Box and whisker plots and normal distributions of impulse responses across decreasing injected column densities for 1-CNT (a,b) and 2-CNT (c,d). The horizontal dashed red line delineates the 3$\sigma$ impulse response. Each column density has a total of 400 iterations with resampled white noise, with the gray boxes representing the inter-quartile range (IQR), the central line showing the median, and the whiskers extending from the minimum to the maximum values in the IQR. Open gray circles indicate statistical outliers. Each data point represents the impulse response of a stack and filter of the top 150 SNR lines in a random white gaussian noise spectrum. The normal distributions on the right are the same distributions as the box and whisker plots and indicate the probability of a 3$\sigma$ or greater response in the top right of each individual panel. The lowest column density with the possibility of a higher than 97\% impulse response is selected for the typical stacking upper limit, which was marked with a red square.}
    \label{fig:stacking_upper_limit}
\end{figure}

To quantify the recovery threshold, we repeated the injection analysis over a grid of decreasing column densities for each isomer. At each column density, 400 independent realizations were generated by resampling the white-noise component while keeping the same observational frequency coverage and local RMS structure. The resulting distributions of peak impulse responses are shown in Figure~\ref{fig:stacking_upper_limit}. The gray boxes summarize the inter-quartile ranges of the recovered impulse responses, while the overplotted points show the individual iterations with randomly resampled noise. The dashed horizontal line marks the $3\sigma$ matched-filter response threshold. For each column density, the impulse-response distribution was also approximated by a normal distribution, and the probability of obtaining a response greater than $3\sigma$ was calculated. We define the stacking upper limit as the lowest injected column density for which the probability of recovering a 3σ response remains greater than 97\%.

Using this criterion, the stacking upper limit for 1-CNT is $3.5 \times 10^{11}~\mathrm{cm^{-2}}$, for which the injected spectra yield a $>$3$\sigma$ matched-filter response in 97.94\% of the trials. For 2-CNT, the corresponding stacking upper limit is $2.5 \times 10^{11}~\mathrm{cm^{-2}}$, with a $>3\sigma$ recovery probability of 97.03\%. These limits are nearly one order of magnitude lower than the single-transition upper limits of $2.18 \times 10^{12}~\mathrm{cm^{-2}}$ for 1-CNT and $1.54 \times 10^{12}~\mathrm{cm^{-2}}$ for 2-CNT, demonstrating the improved sensitivity obtained by combining many weak transitions through stacking and matched filtering.

\begin{table}[htb!]
    \caption{Source parameters used in 1-CNT and 2-CNT molecular simulations based on observations of benzonitrile towards TMC-1\citep{DR5.3}. Each column density of the component is 1/4 of the total column density.} 
    \label{table:sourceparams}
    \centering
    \begin{tabular}{ccccccccc}
    \toprule
      Component &  $v_\mathrm{LSR}$	&	Size	&	$T_\mathrm{ex}$	&	$\Delta V$	\\
	No. & [$\mathrm{km\,s^{-1}}$] &[$^{\prime\prime}$]	&	[$\mathrm{K}$]	&	[$\mathrm{km\,s^{-1}}$]\\
	\midrule
    1 &$5.61$ &$164$ & \multirow{4}{*}{$9.9$} &\multirow{4}{*}{$0.1214$}\\
    2 &$5.78$ &$106$ & &\\
    3 &$5.89$ &$116$ & &\\
	4 &$6.02$ &$130$ & &\\
	\bottomrule
\\
    \end{tabular}
\end{table}

\clearpage

\section{Ab initio calculations of potential formation routes}

All ab initio calculations of reaction pathways for the reaction of CN with triphenylene were performed with ORCA 6.1.0 \citep{UHF:2000,ORCA:2012,SHARK:2023,ORCA:2025}. Reaction pathways were evaluated using a range of density functional theory (DFT) methods: RI-BP86 \citep{Perdew:1986:8822,Becke:1988:3098,Lee:1988:785}, B3LYP \citep{B3LYP,Becke:1988:3098,Lee:1988:785,Stephens:1994:11623}, M06-2X \citep{Zhao:2008:1432}, and $\omega$B97X \citep{Chai:2008:084106}, along with empirical dispersion corrections D3(BJ) for B3LYP and RI-BP86 \citep{D3BJ,Becke:2006:221101}, D3(0) for M06-2X\citep{D3BJ}, and D4 for $\omega$B97X \citep{Caldeweyher:2017:034112,Caldeweyher:2019:154122}. Initial structures for the adducts and separated reagents were optimized with the RI-BP86 DFT functional \citep{Perdew:1986:8822,Becke:1988:3098,Lee:1988:785} using the minimally augmented ma-def2-SVP basis set \citep{def2TZVPP2,Zheng:2011:295}, the def2/J auxiliary basis set \citep{def2TZVPP1}, and D3(BJ) empirical dispersion corrections \citep{D3BJ,Becke:2006:221101}. The presence of barriers to the addition of \ce{CN} to triphenylene and the subsequent elimination of an H atom was evaluated by carrying out relaxed scans of the forming or breaking bonds. In these scans, all coordinates except the bond being scanned were allowed to relax while the scanned coordinate was fixed. Maxima from these scans were used as input geometries for transition-state optimizations, whereas minima were used as input geometries for complexes. Harmonic vibrational analysis was carried out to verify that these structures were indeed stationary points, and the resulting frequencies and harmonic zero-point energies were scaled by 0.9956 and 1.0207, respectively, in the manner suggested by \citet{Kesharwani:2015:1701}. Intrinsic reaction coordinate (IRC) scans were performed to verify that transition states linked reactants and products~\citep{Ishida:1977:2153,orca2020}. The structures found by this approach were reoptimized with the hybrid B3LYP \citep{B3LYP,Becke:1988:3098,Lee:1988:785,Stephens:1994:11623}, M06-2X \citep{Zhao:2008:1432}, and $\omega$B97X functionals \citep{Chai:2008:084106} with the D3(BJ) \citep{D3BJ,Becke:2006:221101}, D3 \citep{D3BJ}, and D4 \citep{Caldeweyher:2017:034112,Caldeweyher:2019:154122} empirical dispersion corrections, respectively, and the minimally augmented ma-def2-TZVPP triple-zeta basis set \citep{def2TZVPP2,Zheng:2011:295}. Harmonic vibrational frequencies were calculated for these structures, and the results and harmonic zero-point energies were scaled using the factors calculated by \citet{Kesharwani:2015:1701} (B3LYP vib 0.96568, zpe 0.98831; M062X vib 0.94986, zpe 0.97483; $\omega$B97X vib 0.95334, zpe 0.97795). In addition, the initial CN addition pathways were reevaluated by relaxed scans of adduct formation using B3LYP-D3(BJ), M06-2X-D3, and $\omega$B97X-D4 with ma-def2-TZVPP, and these scans were used to generate input geometries for the entrance-channel transition states. Single-point energy corrections were carried out using the explicitly correlated domain pair local natural orbital coupled cluster method including single and double excitations and perturbative triples, DLPNO-CCSD(T)-F12, with tight thresholds for the PNOs \citep{Pinski:2015:034108,Riplinger:2016:024109}. This method was used with the modified double-zeta correlation-consistent basis set cc-pVDZ-F12~\citep{Dunning:1989:1007a,Kendall:1992:6796,Davidson:1996:514} optimized for explicitly correlated methods for the orbital basis set. The DLPNO-CCSD(T)-F12 calculations also used the aug-cc-pVTZ/C auxiliary basis sets \citep{Dunning:1989:1007a,Kendall:1992:6796,Weigend:2002:3175,Hatig:2005:59} and cc-pVDZ-F12-CABS near-complete complementary auxiliary basis sets (CABS) \citep{Peterson:2008:084102,Noga:2009:1,Valeev:2004:190} taken from the Basis Set Exchange \citep{Feller:1996:1571,Schuchardt:2007:1045}.  

\begin{figure*}[t!]
\begin{center}
    \includegraphics[width=\linewidth]{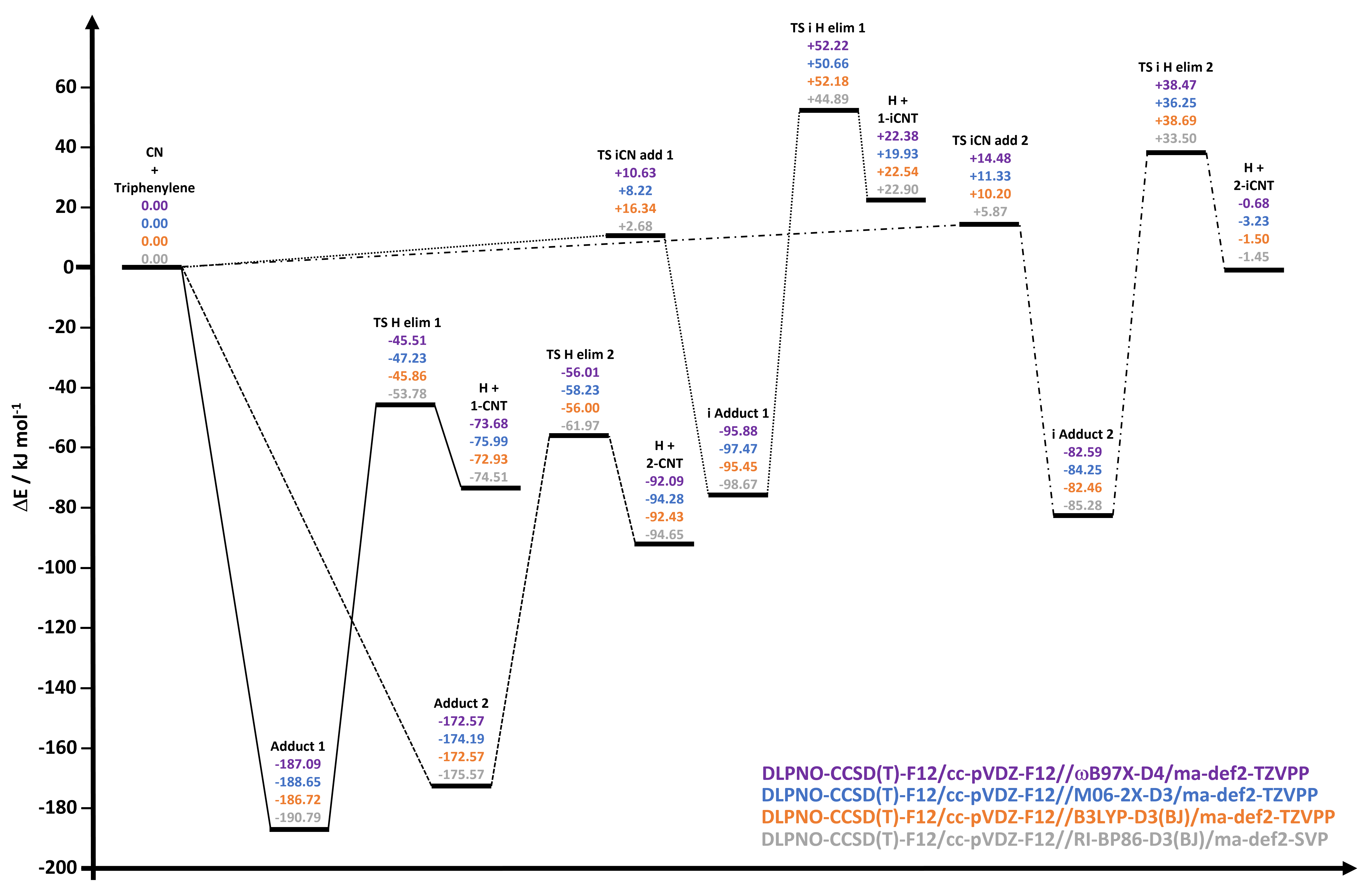}
    \caption{The calculated potential energy surface for the cyano group tagging reaction to form 1-CNT, 2-CNT, 1-iCNT, and 2-iCNT. The computation was carried out by DLPNO-CCSD(T)-F12/cc-pVDZ-F12//several different DFT levels.}
    \label{fig:PES}
\end{center}
\end{figure*}

When considering the presence of entrance-channel complexes, none of the hybrid functionals optimized loose entrance-channel complexes; instead, they resulted in the formation of tightly bound adducts. Additionally, where the RI-BP86-D3(BJ) functional did successfully optimize structures in the entrance channel that appeared to be loose complexes, inclusion of DLPNO-CCSD(T)-F12/cc-pVDZ-F12 single-point energy corrections caused these minima to disappear. The fact that loose complexes found with RI-BP86-D3(BJ) optimize to other adducts with the hybrid functionals suggests that the assigned entrance-channel barriers may not actually connect the entrance channel to the adducts, but may instead represent barriers to conversion between adducts. It is nevertheless apparent that the formation of the isocyano adducts occurs over emerged barriers ($>$ 10 kJ mol$^{-1}$), with the subsequent H atom elimination proceeding over even higher emerged barriers leading to the formation of isocyanotriphenylene. As a result, we conclude that the reaction of CN with triphenylene will lead only to the cyano-triphenylenes at low temperatures. The formation of both 1- and 2-cyanotriphenylene is exothermic and proceeds over very submerged entrance-channel (CN addition) and exit (H atom elimination) barriers. Therefore, under kinetic control at low temperatures both 1- and 2-cyanotriphenylene are likely to form. If the reaction is instead under thermodynamic control, the energy difference between 1- and 2-cyanotriphenylene ($>$ 15 kJ mol$^{-1}$) is sufficient to support predominantly the formation of 2-cyanotriphenylene. Interestingly, the 1-CNT addition intermediate (adduct 1) is lower in energy than its 2-CNT counterpart (adduct 2). This difference can be attributed to the partial relief of steric strain through the out-of-plane displacement of the cyano group in adduct 1, whereas adduct 2 retains steric repulsion between the hydrogen atoms in all three bay regions.

\begin{table}
    \centering
    \begin{tabular}{ccccc}
 & \multicolumn{4}{c}{DLPNO-CCSD(T)-F12/cc-pVDZ-F12//X (kJ mol$^{-1}$)}\\
         &  RI-BP86-D3(BJ)&  B3LYP-D3(BJ)&  M06-2X-D3& $\omega$B97X-D4\\
         CN + Triphenylene&  0&  0&  0& 0\\
         Complex 1&  33.93&  ...&  ...& ...\\
         Complex 2&  49.25
&  ...&  ...& ...\\
         Addition TS 1 &  -22.86
&  -21.86&  -25.80
& -23.26
\\
         Adduct 1&  -190.79&  -188.72&  -188.65& -187.09
\\
         H elim TS 1&  -53.78&  -45.86&  -47.23& -45.51
\\
         H + 1-CNT&  -74.51
&  -72.95&  -75.99& -73.68
\\
         Addition TS 2&  -22.96
&  -22.49&  -25.38& -22.77
\\
         Adduct 2&  -175.57
&  -172.75&  -174.19& -172.57
\\
 H elim TS 2& -61.97
& -56.00& -58.23&-56.01
\\
 H + 2-CNT& -94.65
& -92.43& -94.28&-92.09
\\
 Addition TS i1& 2.68& 16.34& 8.22&10.63
\\
 Adduct i1& -98.67& -95.45& -97.47&-95.88
\\
 H elim TS i1& 44.89& 52.18& 50.66&52.22
\\
 H + 1-iCNT& 22.90& 22.54& 19.93&22.38
\\
 Addition TS i2& 5.87
& 10.20& 11.33&14.48
\\
 Adduct i2& -85.28
& -82.46& -84.25
&-82.59
\\
 H elim TS i2& 33.50
& 38.69
& 36.25
&
38.47\\
 H + 2-iCNT& -1.45
& -1.50& -3.23&-0.68
\\
    \end{tabular}
    \caption{Relative energies along the calculated \ce{CN + triphenylene} reaction pathways, evaluated at the DLPNO-CCSD(T)-F12/cc-pVDZ-F12//X levels of theory and reported in kJ mol$^{-1}$ relative to the separated \ce{CN + triphenylene} reactants.}
    \label{tab:placeholder}
\end{table}

\section {Partition Functions of Cyanotriphenylenes}
\label{app:qpart}
The partition functions calculated by SPCAT for cyanotriphenylenes are
reported in Table~\ref{tab:qrot}.

\begin{table}[htb!]
    \centering
    \caption{Partition functions for cyanotriphenylenes used in the upper limit calculation and stacking analysis. The values were calculated with SPCAT and \texttt{molsim} \citep{molsim} using $J_\mathrm{max}=400$ and $K_\mathrm{max}=350$.}
    \label{tab:qrot}
    {
    \begin{tabular}{rrr}
    \toprule
        {Temperature [K]} & {1-CNT} & {2-CNT} \\
    \midrule
        1.0   & 2638.0034     & 3052.5820     \\
        3.0   & 13676.8597    & 15830.3224    \\
        5.0   & 29414.8579    & 34048.0255    \\
        7.0   & 48716.6126    & 56391.1946    \\
        9.375 & 75498.1010    & 87392.8008    \\
        12.0  & 109324.8757   & 126549.9036   \\
        15.0  & 152778.4215   & 176850.7156   \\
        18.75 & 213506.1898   & 247147.6510   \\
        25.0  & 328703.1822   & 380496.3184   \\
        37.5  & 603852.5198   & 698998.5361   \\
        45.0  & 793783.3439   & 918853.0663   \\
        55.0  & 1072578.1977  & 1241569.4208  \\
        65.0  & 1378029.4400  & 1595138.0907  \\
        75.0  & 1707988.6210  & 1977071.5394  \\
        95.0  & 2434947.5068  & 2818525.7562  \\
        115.0 & 3243122.8428  & 3753965.0585  \\
        135.0 & 4125055.5442  & 4774729.9527  \\
        150.0 & 4831411.5360  & 5592206.7243  \\
        175.0 & 6088335.3868  & 7046520.5891  \\
        200.0 & 7438143.8358  & 8607371.4988  \\
        225.0 & 8874014.7101  & 10265940.1020 \\
        250.0 & 10389647.4284 & 12013717.8320 \\
        275.0 & 11978942.5944 & 13842248.0074 \\
        300.0 & 13635838.0778 & 15743065.0078 \\
        400.0 & 20817804.8544 & 23903283.0698 \\
        500.0 & 28604243.3686 & 32592167.1279 \\
    \bottomrule
    \end{tabular}
    }
\end{table}

\clearpage
\bibliographystyle{aasjournalv7}
\bibliography{references,si_refs}

@article{PAHsincarbon,
  title = {Detection and {{Characterization}} of {{Cold Interstellar Dust}} and {{Polycyclic Aromatic Hydrocarbon Emission}}, from {{COBE Observations}}},
  author = {Dwek, E. and Arendt, R. G. and Fixsen, D. J. and Sodroski, T. J. and Odegard, N. and Weiland, J. L. and Reach, W. T. and Hauser, M. G. and Kelsall, T. and Moseley, S. H. and Silverberg, R. F. and Shafer, R. A. and Ballester, J. and Bazell, D. and Isaacman, R.},
  year = 1997,
  month = feb,
  journal = {The Astrophysical Journal},
  volume = {475},
  pages = {565--579},
  publisher = {IOP},
  issn = {0004-637X},
  doi = {10.1086/303568},
  urldate = {2026-03-12},
}

@article{PAHsincarbon2,
  title = {Interstellar {{Polycyclic Aromatic Hydrocarbon Molecules}}},
  author = {Tielens, A.G.G.M.},
  year = 2008,
  month = sep,
  journal = {Annual Review of Astronomy and Astrophysics},
  volume = {46},
  number = {1},
  pages = {289--337},
  issn = {0066-4146, 1545-4282},
  doi = {10.1146/annurev.astro.46.060407.145211},
  urldate = {2026-03-12},
  langid = {english}
}

@article{BrettMcCarthyReview,
  title = {Spectroscopic {{Studies}} of {{Polycyclic Aromatic Hydrocarbons}}: {{Interstellar Aromatic Chemistry Revealed}}},
  shorttitle = {Spectroscopic {{Studies}} of {{Polycyclic Aromatic Hydrocarbons}}},
  author = {McCarthy, Michael C. and McGuire, Brett A.},
  year = 2026,
  journal = {Annual Review of Physical Chemistry},
  volume = {77},
  pages = {17.1-17.25},
  publisher = {Annual Reviews},
  doi = {10.1146/annurev-physchem-082324-010544},
  urldate = {2026-03-12},
  langid = {english}
}

@article{Tweety,
  title = {Laboratory {{Rotational Spectra}} of {{Cyanocyclohexane}} and {{Its Siblings}} (1- and 4-{{Cyanocyclohexene}}) {{Using}} a {{Compact CP-FTMW Spectrometer}} for {{Interstellar Detection}}},
  author = {Wenzel, Gabi and Holdren, Martin S. and Stewart, D. Archie and Toru Shay, Hannah and Byrne, Alex N. and Xue, Ci and McGuire, Brett A.},
  year = {2025},
  month = may,
  journal = {The Journal of Physical Chemistry A},
  volume = {129},
  number = {18},
  pages = {3986--4001},
  issn = {1089-5639, 1520-5215},
  doi = {10.1021/acs.jpca.5c00980},
  urldate = {2025-05-14},
}

@article{cyanopyrene2,
  title = {Detections of Interstellar Aromatic Nitriles 2-Cyanopyrene and 4-Cyanopyrene in {{TMC-1}}},
  author = {Wenzel, Gabi and Speak, Thomas H. and Changala, P. Bryan and Willis, Reace H. J. and Burkhardt, Andrew M. and Zhang, Shuo and Bergin, Edwin A. and Byrne, Alex N. and Charnley, Steven B. and Fried, Zachary T. P. and Gupta, Harshal and Herbst, Eric and Holdren, Martin S. and Lipnicky, Andrew and Loomis, Ryan A. and Shingledecker, Christopher N. and Xue, Ci and Remijan, Anthony J. and Wendlandt, Alison E. and McCarthy, Michael C. and Cooke, Ilsa R. and McGuire, Brett A.},
  year = 2025,
  month = feb,
  journal = {Nature Astronomy},
  volume = {9},
  number = {2},
  pages = {262--270},
  publisher = {Nature Publishing Group},
  issn = {2397-3366},
  doi = {10.1038/s41550-024-02410-9},
  urldate = {2025-11-20},
  copyright = {2024 The Author(s)},
  langid = {english}
}

@article{NMR1,
  title = {Palladium {{Catalyzed C}}--{{I}} and {{Vicinal C}}--{{H Dual Activation}} of {{Diaryliodonium Salts}} for {{Diarylations}}: {{Synthesis}} of {{Triphenylenes}}},
  shorttitle = {Palladium {{Catalyzed C}}--{{I}} and {{Vicinal C}}--{{H Dual Activation}} of {{Diaryliodonium Salts}} for {{Diarylations}}},
  author = {Wu, Xunshen and Han, Jianwei and Wang, Limin},
  year = 2018,
  month = jan,
  journal = {The Journal of Organic Chemistry},
  volume = {83},
  number = {1},
  pages = {49--56},
  publisher = {American Chemical Society},
  issn = {0022-3263},
  doi = {10.1021/acs.joc.7b01905},
  urldate = {2025-12-18}
}

@article{NMR2,
  title = {Synthesis of Triphenylene Derivatives by {{Pd-catalyzed Suzuki}} Coupling/Intramolecular {{CH}} Activation between Arylboronic Acids and Dibromobiphenyls},
  author = {Tu, Jingxuan and Li, Gaoqiang and Zhao, Xiaoqian and Xu, Feng},
  year = 2019,
  month = jan,
  journal = {Tetrahedron Letters},
  volume = {60},
  number = {1},
  pages = {44--47},
  issn = {0040-4039},
  doi = {10.1016/j.tetlet.2018.11.052},
  urldate = {2025-12-18}
}

@article{cyanonaphthalene,
  title = {Detection of Two Interstellar Polycyclic Aromatic Hydrocarbons via Spectral Matched Filtering},
  author = {McGuire, Brett A. and Loomis, Ryan A. and Burkhardt, Andrew M. and Lee, Kin Long Kelvin and Shingledecker, Christopher N. and Charnley, Steven B. and Cooke, Ilsa R. and Cordiner, Martin A. and Herbst, Eric and Kalenskii, Sergei and Siebert, Mark A. and Willis, Eric R. and Xue, Ci and Remijan, Anthony J. and McCarthy, Michael C.},
  year = {2021},
  month = mar,
  journal = {Science},
  volume = {371},
  number = {6535},
  pages = {1265--1269},
  issn = {0036-8075, 1095-9203},
  doi = {10.1126/science.abb7535},
  urldate = {2025-04-19},
}

@article{indeneB,
  title = {Discovery of the {{Pure Polycyclic Aromatic Hydrocarbon Indene}} (c-{{C9H8}}) with {{GOTHAM Observations}} of {{TMC-1}}},
  author = {Burkhardt, Andrew M. and Long Kelvin Lee, Kin and Bryan Changala, P. and Shingledecker, Christopher N. and Cooke, Ilsa R. and Loomis, Ryan A. and Wei, Hongji and Charnley, Steven B. and Herbst, Eric and McCarthy, Michael C. and McGuire, Brett A.},
  year = {2021},
  month = jun,
  journal = {The Astrophysical Journal Letters},
  volume = {913},
  number = {2},
  pages = {L18},
  issn = {2041-8205, 2041-8213},
  doi = {10.3847/2041-8213/abfd3a},
  urldate = {2025-04-17},
}

@article{indeneC,
  title = {Pure Hydrocarbon Cycles in {{TMC-1}}: {{Discovery}} of Ethynyl Cyclopropenylidene, Cyclopentadiene, and Indene},
  shorttitle = {Pure Hydrocarbon Cycles in {{TMC-1}}},
  author = {Cernicharo, J. and Ag{\'u}ndez, M. and Cabezas, C. and Tercero, B. and Marcelino, N. and Pardo, J. R. and De Vicente, P.},
  year = {2021},
  month = may,
  journal = {Astronomy \& Astrophysics},
  volume = {649},
  pages = {L15},
  issn = {0004-6361, 1432-0746},
  doi = {10.1051/0004-6361/202141156},
  urldate = {2025-04-17},
}

@article{cyanoindene,
  title = {Discovery of {{Interstellar}} 2-{{Cyanoindene}} (2-{{C}}{\textsubscript{9}} {{H}}{\textsubscript{7}} {{CN}}) in {{GOTHAM Observations}} of {{TMC-1}}},
  author = {Sita, Madelyn L. and Changala, P. Bryan and Xue, Ci and Burkhardt, Andrew M. and Shingledecker, Christopher N. and Kelvin Lee, Kin Long and Loomis, Ryan A. and Momjian, Emmanuel and Siebert, Mark A. and Gupta, Divita and Herbst, Eric and Remijan, Anthony J. and McCarthy, Michael C. and Cooke, Ilsa R. and McGuire, Brett A.},
  year = {2022},
  month = oct,
  journal = {The Astrophysical Journal Letters},
  volume = {938},
  number = {2},
  pages = {L12},
  issn = {2041-8205, 2041-8213},
  doi = {10.3847/2041-8213/ac92f4},
  urldate = {2025-01-13},
}

@article{cyanoacenaphthylene,
  title = {Discovery of Two Cyano Derivatives of Acenaphthylene ({{C}}{\textsubscript{12}} {{H}}{\textsubscript{8}} ) in {{TMC-1}} with the {{QUIJOTE}} Line Survey},
  author = {Cernicharo, J. and Cabezas, C. and Fuentetaja, R. and Ag{\'u}ndez, M. and Tercero, B. and Janeiro, J. and Juanes, M. and Kaiser, R. I. and Endo, Y. and Steber, A. L. and P{\'e}rez, D. and P{\'e}rez, C. and Lesarri, A. and Marcelino, N. and De Vicente, P.},
  year = {2024},
  month = oct,
  journal = {Astronomy \& Astrophysics},
  volume = {690},
  pages = {L13},
  issn = {0004-6361, 1432-0746},
  doi = {10.1051/0004-6361/202452196},
  urldate = {2025-04-17},
}

@article{Fluorene,
  title = {Rotational {{Spectra}} of {{Small PAHs}}: {{Acenaphthene}}, {{Acenaphthylene}}, {{Azulene}}, and {{Fluorene}}},
  shorttitle = {Rotational {{Spectra}} of {{Small PAHs}}},
  author = {Thorwirth, S. and Theule, P. and Gottlieb, C. A. and McCarthy, M. C. and Thaddeus, P.},
  year = 2007,
  month = jun,
  journal = {The Astrophysical Journal},
  volume = {662},
  number = {2},
  pages = {1309--1314},
  issn = {0004-637X, 1538-4357},
  doi = {10.1086/518026},
  urldate = {2025-04-17},
  langid = {english}
}

@article{cyanofluorene,
  title = {Rotational Spectra of Five Cyano Derivatives of Fluorene},
  author = {Cabezas, Carlos and Janeiro, Jes{\'u}s and Steber, Amanda L. and P{\'e}rez, Dolores and Berm{\'u}dez, Celina and Guiti{\'a}n, Enrique and Lesarri, Alberto and Cernicharo, Jos{\'e}},
  year = 2024,
  journal = {Physical Chemistry Chemical Physics},
  volume = {26},
  number = {36},
  pages = {23703--23709},
  issn = {1463-9076, 1463-9084},
  doi = {10.1039/D4CP01924E},
  urldate = {2025-01-06},
  langid = {english}
}

@article{iminofluorene,
  title = {Rotational {{Spectra}} and {{Search}} for {{Aromatic Imines}}: 9-Iminofluorene and {{Benzophenone Imine}}},
  shorttitle = {Rotational {{Spectra}} and {{Search}} for {{Aromatic Imines}}},
  author = {Ren, Huanyu and Stewart, D. Archie and Wenzel, Gabi and Speak, Thomas H. and Holdren, Martin S. and Willis, Reace H. J. and McGuire, Brett A.},
  year = 2026,
  month = Jan,
  journal = {The Astrophysical Journal},
  volume = {996},
  number = {1},
  pages = {108},
  publisher = {The American Astronomical Society},
  issn = {0004-637X},
  doi = {10.3847/1538-4357/ae29ab},
  urldate = {2026-01-06},
  langid = {english}
}

@article{cyanocoronene,
  title = {Discovery of the {{Seven-ring Polycyclic Aromatic Hydrocarbon Cyanocoronene}} ({{C24H11CN}}) in {{GOTHAM Observations}} of {{TMC-1}}},
  author = {Wenzel, Gabi and Gong, Siyuan and Xue, Ci and Changala, P. Bryan and Holdren, Martin S. and Speak, Thomas H. and Stewart, D. Archie and Fried, Zachary T. P. and Willis, Reace H. J. and Bergin, Edwin A. and Burkhardt, Andrew M. and Byrne, Alex N. and Charnley, Steven B. and Lipnicky, Andrew and Loomis, Ryan A. and Shingledecker, Christopher N. and Cooke, Ilsa R. and McCarthy, Michael C. and Remijan, Anthony J. and Wendlandt, Alison E. and McGuire, Brett A.},
  year = {2025},
  month = apr,
  journal = {The Astrophysical Journal Letters},
  shortjournal = {ApJL},
  volume = {984},
  number = {1},
  pages = {L36},
  publisher = {The American Astronomical Society},
  issn = {2041-8205},
  doi = {10.3847/2041-8213/adc911},
  url = {https://dx.doi.org/10.3847/2041-8213/adc911},
  urldate = {2025-05-27},
}

@article{cyanopyrene,
  title = {Detection of Interstellar 1-Cyanopyrene: {{A}} Four-Ring Polycyclic Aromatic Hydrocarbon},
  shorttitle = {Detection of Interstellar 1-Cyanopyrene},
  author = {Wenzel, Gabi and Cooke, Ilsa R. and Changala, P. Bryan and Bergin, Edwin A. and Zhang, Shuo and Burkhardt, Andrew M. and Byrne, Alex N. and Charnley, Steven B. and Cordiner, Martin A. and Duffy, Miya and Fried, Zachary T. P. and Gupta, Harshal and Holdren, Martin S. and Lipnicky, Andrew and Loomis, Ryan A. and Shay, Hannah Toru and Shingledecker, Christopher N. and Siebert, Mark A. and Stewart, D. Archie and Willis, Reace H. J. and Xue, Ci and Remijan, Anthony J. and Wendlandt, Alison E. and McCarthy, Michael C. and McGuire, Brett A.},
  year = {2024},
  month = nov,
  journal = {Science},
  volume = {386},
  number = {6723},
  pages = {810--813},
  publisher = {American Association for the Advancement of Science},
  doi = {10.1126/science.adq6391},
  urldate = {2024-12-24},
}

@article{cyanophenanthrene,
  title = {Laboratory Rotational Spectroscopy of Cyano Substituted Polycyclic Aromatic Hydrocarbons},
  author = {McNaughton, Don and Jahn, Michaela K and Travers, Michael J and Wachsmuth, Dennis and Godfrey, Peter D and Grabow, Jens-Uwe},
  year = 2018,
  month = jun,
  journal = {Monthly Notices of the Royal Astronomical Society},
  volume = {476},
  number = {4},
  pages = {5268--5273},
  issn = {0035-8711, 1365-2966},
  doi = {10.1093/mnras/sty557},
  urldate = {2026-01-06},
  langid = {english}
}

@article{phenalene,
  title = {Discovery of Interstellar Phenalene (c-{{C13H10}}): {{A}} New Piece in the Chemical Puzzle of {{PAHs}} in Space},
  shorttitle = {Discovery of Interstellar Phenalene (c-{{C13H10}})},
  author = {Cabezas, C. and Ag{\'u}ndez, M. and P{\'e}rez, C. and {Villar-Castro}, D. and Molpeceres, G. and P{\'e}rez, D. and Steber, A. L. and Fuentetaja, R. and Tercero, B. and Marcelino, N. and Lesarri, A. and de Vicente, P. and Cernicharo, J.},
  year = 2025,
  month = sep,
  journal = {Astronomy \& Astrophysics},
  volume = {701},
  pages = {L8},
  publisher = {EDP Sciences},
  issn = {0004-6361, 1432-0746},
  doi = {10.1051/0004-6361/202556687},
  urldate = {2025-11-03},
  copyright = {\copyright{} The Authors 2025},
  langid = {english}
}

@article{Clar,
  title = {Forty Years of {{Clar}}'s Aromatic {$\pi$}-Sextet Rule},
  author = {Sol{\`a}, Miquel},
  year = 2013,
  month = oct,
  journal = {Frontiers in Chemistry},
  volume = {1},
  pages = {22},
  issn = {2296-2646},
  doi = {10.3389/fchem.2013.00022},
  urldate = {2026-01-06},
  pmcid = {PMC3982536},
  pmid = {24790950}
}

@article{inert,
  title = {Towards {{Graphite}}: {{Magnetic Properties}} of {{Large Polybenzenoid Hydrocarbons}}},
  shorttitle = {Towards {{Graphite}}},
  author = {Moran, Damian and Stahl, Frank and Bettinger, Holger F. and Schaefer, Henry F. and Schleyer, Paul V. R.},
  year = 2003,
  month = jun,
  journal = {Journal of the American Chemical Society},
  volume = {125},
  number = {22},
  pages = {6746--6752},
  issn = {0002-7863, 1520-5126},
  doi = {10.1021/ja034497z},
  urldate = {2026-01-06},
  langid = {english}
}

@article{C60,
  title = {Detection of {{C60}} and {{C70}} in a {{Young Planetary Nebula}}},
  author = {Cami, Jan and {Bernard-Salas}, Jeronimo and Peeters, Els and Malek, Sarah Elizabeth},
  year = 2010,
  month = sep,
  journal = {Science},
  volume = {329},
  number = {5996},
  pages = {1180--1182},
  publisher = {American Association for the Advancement of Science},
  doi = {10.1126/science.1192035},
  urldate = {2026-01-06}
}

@article{C60+1,
  title = {Detection of Two Interstellar Absorption Bands Coincident with Spectral Features of {{C60}}+},
  author = {Foing, B. H. and Ehrenfreund, P.},
  year = 1994,
  month = may,
  journal = {Nature},
  volume = {369},
  number = {6478},
  pages = {296--298},
  publisher = {Nature Publishing Group},
  issn = {1476-4687},
  doi = {10.1038/369296a0},
  urldate = {2026-01-06},
  copyright = {1994 Springer Nature Limited},
  langid = {english}
}

@article{C60+2,
  title = {Laboratory Confirmation of {{C60}}+ as the Carrier of Two Diffuse Interstellar Bands},
  author = {Campbell, E. K. and Holz, M. and Gerlich, D. and Maier, J. P.},
  year = 2015,
  month = jul,
  journal = {Nature},
  volume = {523},
  number = {7560},
  pages = {322--323},
  publisher = {Nature Publishing Group},
  issn = {1476-4687},
  doi = {10.1038/nature14566},
  urldate = {2026-01-06},
  copyright = {2015 Springer Nature Limited},
  langid = {english}
}

@article{C60+3,
  title = {Confirming {{Interstellar C60}}+ {{Using}} the {{Hubble Space Telescope}}},
  author = {Cordiner, M. A. and Linnartz, H. and Cox, N. L. J. and Cami, J. and Najarro, F. and Proffitt, C. R. and Lallement, R. and Ehrenfreund, P. and Foing, B. H. and Gull, T. R. and Sarre, P. J. and Charnley, S. B.},
  year = 2019,
  month = apr,
  journal = {The Astrophysical Journal Letters},
  volume = {875},
  number = {2},
  pages = {L28},
  publisher = {The American Astronomical Society},
  issn = {2041-8205},
  doi = {10.3847/2041-8213/ab14e5},
  urldate = {2026-01-06},
  langid = {english}
}

@article{meteorite,
  title = {Isomer Discrimination of Polycyclic Aromatic Hydrocarbons in the {{Murchison}} Meteorite by Resonant Ionization},
  author = {Callahan, Michael P. and {Abo-Riziq}, Ali and Crews, Bridgit and Grace, Louis and {de Vries}, Mattanjah S.},
  year = 2008,
  month = dec,
  journal = {Spectrochimica Acta Part A: Molecular and Biomolecular Spectroscopy},
  volume = {71},
  number = {4},
  pages = {1492--1495},
  issn = {1386-1425},
  doi = {10.1016/j.saa.2008.05.005},
  urldate = {2026-01-07}
}

@article{Bennu,
  title = {Prebiotic Organic Compounds in Samples of Asteroid {{Bennu}} Indicate Heterogeneous Aqueous Alteration},
  author = {Mojarro, Angel and Aponte, Jos{\'e} C. and Dworkin, Jason P. and Elsila, Jamie E. and Glavin, Daniel P. and Connolly, Harold C. and Lauretta, Dante S.},
  year = 2025,
  month = dec,
  journal = {Proceedings of the National Academy of Sciences},
  volume = {122},
  number = {49},
  pages = {e2512461122},
  publisher = {Proceedings of the National Academy of Sciences},
  doi = {10.1073/pnas.2512461122},
  urldate = {2026-01-07}
}

@article{Ryugu,
  title = {Soluble Organic Molecules in Samples of the Carbonaceous Asteroid (162173) {{Ryugu}}},
  author = {Naraoka, Hiroshi and Takano, Yoshinori and Dworkin, Jason P. and Oba, Yasuhiro and Hamase, Kenji and Furusho, Aogu and Ogawa, Nanako O. and Hashiguchi, Minako and Fukushima, Kazuhiko and Aoki, Dan and {Schmitt-Kopplin}, Philippe and Aponte, Jos{\'e} C. and Parker, Eric T. and Glavin, Daniel P. and McLain, Hannah L. and Elsila, Jamie E. and Graham, Heather V. and Eiler, John M. and {Orthous-Daunay}, Francois-Regis and Wolters, C{\'e}dric and Isa, Junko and Vuitton, V{\'e}ronique and Thissen, Roland and Sakai, Saburo and Yoshimura, Toshihiro and Koga, Toshiki and Ohkouchi, Naohiko and Chikaraishi, Yoshito and Sugahara, Haruna and Mita, Hajime and Furukawa, Yoshihiro and Hertkorn, Norbert and Ruf, Alexander and Yurimoto, Hisayoshi and Nakamura, Tomoki and Noguchi, Takaaki and Okazaki, Ryuji and Yabuta, Hikaru and Sakamoto, Kanako and Tachibana, Shogo and Connolly, Harold C. and Lauretta, Dante S. and Abe, Masanao and Yada, Toru and Nishimura, Masahiro and Yogata, Kasumi and Nakato, Aiko and Yoshitake, Miwa and Suzuki, Ayako and Miyazaki, Akiko and Furuya, Shizuho and Hatakeda, Kentaro and Soejima, Hiromichi and Hitomi, Yuya and Kumagai, Kazuya and Usui, Tomohiro and Hayashi, Tasuku and Yamamoto, Daiki and Fukai, Ryota and Kitazato, Kohei and Sugita, Seiji and Namiki, Noriyuki and Arakawa, Masahiko and Ikeda, Hitoshi and Ishiguro, Masateru and Hirata, Naru and Wada, Koji and Ishihara, Yoshiaki and Noguchi, Rina and Morota, Tomokatsu and Sakatani, Naoya and Matsumoto, Koji and Senshu, Hiroki and Honda, Rie and Tatsumi, Eri and Yokota, Yasuhiro and Honda, Chikatoshi and Michikami, Tatsuhiro and Matsuoka, Moe and Miura, Akira and Noda, Hirotomo and Yamada, Tetsuya and Yoshihara, Keisuke and Kawahara, Kosuke and Ozaki, Masanobu and Iijima, Yu-ichi and Yano, Hajime and Hayakawa, Masahiko and Iwata, Takahiro and Tsukizaki, Ryudo and Sawada, Hirotaka and Hosoda, Satoshi and Ogawa, Kazunori and Okamoto, Chisato and Hirata, Naoyuki and Shirai, Kei and Shimaki, Yuri and Yamada, Manabu and Okada, Tatsuaki and Yamamoto, Yukio and Takeuchi, Hiroshi and Fujii, Atsushi and Takei, Yuto and Yoshikawa, Kento and Mimasu, Yuya and Ono, Go and Ogawa, Naoko and Kikuchi, Shota and Nakazawa, Satoru and Terui, Fuyuto and Tanaka, Satoshi and Saiki, Takanao and Yoshikawa, Makoto and Watanabe, Sei-ichiro and Tsuda, Yuichi},
  year = 2023,
  month = feb,
  journal = {Science},
  volume = {379},
  number = {6634},
  pages = {eabn9033},
  publisher = {American Association for the Advancement of Science},
  doi = {10.1126/science.abn9033},
  urldate = {2026-01-07}
}

@article{gasphase,
  title = {Gas-{{Phase Synthesis}} of {{Triphenylene}} ({{C18H12}})},
  author = {Zhao, Long and Xu, Bo and Ablikim, Utuq and Lu, Wenchao and Ahmed, Musahid and Evseev, Mikhail M. and Bashkirov, Eugene K. and Azyazov, Valeriy N. and Howlader, A. Hasan and Wnuk, Stanislaw F. and Mebel, Alexander M. and Kaiser, Ralf I.},
  year = 2019,
  journal = {ChemPhysChem},
  volume = {20},
  number = {6},
  pages = {791--797},
  issn = {1439-7641},
  doi = {10.1002/cphc.201801154},
  urldate = {2025-12-19},
  copyright = {\copyright{} 2019 Wiley-VCH Verlag GmbH \& Co. KGaA, Weinheim},
  langid = {english}
}

@article{DR5.3,
  title = {The {{Molecular Inventory}} of {{TMC-1}} with {{GOTHAM Observations}}},
  author = {Xue, Ci and Byrne, Alex N. and Morgan, Larry and Wenzel, Gabi and Changala, P. Bryan and Fried, Zachary T. P. and Loomis, Ryan A. and Remijan, Anthony and Bergin, Edwin A. and Cooke, Ilsa R. and Frayer, David and Burkhardt, Andrew M. and Charnley, Steven B. and Cordiner, Martin A. and Lipnicky, Andrew and McCarthy, Michael C. and McGuire, Brett A.},
  year = 2025,
  month = nov,
  journal = {The Astrophysical Journal Supplement Series},
  volume = {281},
  number = {1},
  pages = {9},
  issn = {0067-0049, 1538-4365},
  doi = {10.3847/1538-4365/ae04e5},
  urldate = {2025-11-20},
  langid = {english}
}

@article{benzyne,
  title = {Discovery of Benzyne, {\emph{o}} -{{C}}{\textsubscript{6}} {{H}}{\textsubscript{4}} , in {{TMC-1}} with the {{QUIJOTE}} Line Survey},
  author = {Cernicharo, J. and Ag{\'u}ndez, M. and Kaiser, R. I. and Cabezas, C. and Tercero, B. and Marcelino, N. and Pardo, J. R. and De Vicente, P.},
  year = 2021,
  month = aug,
  journal = {Astronomy \& Astrophysics},
  volume = {652},
  pages = {L9},
  issn = {0004-6361, 1432-0746},
  doi = {10.1051/0004-6361/202141660},
  urldate = {2026-01-11},
  copyright = {https://www.edpsciences.org/en/authors/copyright-and-licensing},
  langid = {english}
}

@misc{g16c02,
  title = {Gaussian 16 {{Revision C}}.02},
  author = {Frisch, M. J. and Trucks, G. W. and Schlegel, H. B. and Scuseria, G. E. and Robb, M. A. and Cheeseman, J. R. and Scalmani, G. and Barone, V. and Petersson, G. A. and Nakatsuji, H. and Li, X. and Caricato, M. and Marenich, A. V. and Bloino, J. and Janesko, B. G. and Gomperts, R. and Mennucci, B. and Hratchian, H. P. and Ortiz, J. V. and Izmaylov, A. F. and Sonnenberg, J. L. and {Williams-Young}, D. and Ding, F. and Lipparini, F. and Egidi, F. and Goings, J. and Peng, B. and Petrone, A. and Henderson, T. and Ranasinghe, D. and Zakrzewski, V. G. and Gao, J. and Rega, N. and Zheng, G. and Liang, W. and Hada, M. and Ehara, M. and Toyota, K. and Fukuda, R. and Hasegawa, J. and Ishida, M. and Nakajima, T. and Honda, Y. and Kitao, O. and Nakai, H. and Vreven, T. and Throssell, K. and Montgomery, Jr., J. A. and Peralta, J. E. and Ogliaro, F. and Bearpark, M. J. and Heyd, J. J. and Brothers, E. N. and Kudin, K. N. and Staroverov, V. N. and Keith, T. A. and Kobayashi, R. and Normand, J. and Raghavachari, K. and Rendell, A. P. and Burant, J. C. and Iyengar, S. S. and Tomasi, J. and Cossi, M. and Millam, J. M. and Klene, M. and Adamo, C. and Cammi, R. and Ochterski, J. W. and Martin, R. L. and Morokuma, K. and Farkas, O. and Foresman, J. B. and Fox, D. J.},
  year = {2019}
}

@article{6-311g,
  title = {Self-consistent Molecular Orbital Methods. {{XX}}. {{A}} Basis Set for Correlated Wave Functions},
  author = {Krishnan, R. and Binkley, J. S. and Seeger, R. and Pople, J. A.},
  year = {1980},
  month = jan,
  journal = {The Journal of Chemical Physics},
  volume = {72},
  number = {1},
  pages = {650--654},
  issn = {0021-9606},
  doi = {10.1063/1.438955},
  urldate = {2025-08-12}
}

@article{B3LYP,
  title = {Density-functional Thermochemistry. {{III}}. {{The}} Role of Exact Exchange},
  author = {Becke, Axel D.},
  year = {1993},
  month = apr,
  journal = {The Journal of Chemical Physics},
  volume = {98},
  number = {7},
  pages = {5648--5652},
  issn = {0021-9606},
  doi = {10.1063/1.464913},
  urldate = {2025-08-12}
}

@article{D3BJ,
  title = {Effect of the Damping Function in Dispersion Corrected Density Functional Theory},
  author = {Grimme, Stefan and Ehrlich, Stephan and Goerigk, Lars},
  year = {2011},
  journal = {Journal of Computational Chemistry},
  volume = {32},
  number = {7},
  pages = {1456--1465},
  issn = {1096-987X},
  doi = {10.1002/jcc.21759},
  urldate = {2025-08-12},
  copyright = {Copyright {\copyright} 2011 Wiley Periodicals, Inc.},
  langid = {english}
}

@article{spcat,
  title = {The Fitting and Prediction of Vibration-Rotation Spectra with Spin Interactions},
  author = {Pickett, Herbert M.},
  year = {1991},
  journal = {Journal of Molecular Spectroscopy},
  volume = {148},
  number = {2},
  pages = {371--377},
  issn = {0022-2852},
  doi = {10.1016/0022-2852(91)90393-O},
}

@article{GOTHAM,
  title = {Early {{Science}} from {{GOTHAM}}: {{Project Overview}}, {{Methods}}, and the {{Detection}} of {{Interstellar Propargyl Cyanide}} ({{HCCCH2CN}}) in {{TMC-1}}},
  shorttitle = {Early {{Science}} from {{GOTHAM}}},
  author = {McGuire, Brett A. and Burkhardt, Andrew M. and Loomis, Ryan A. and Shingledecker, Christopher N. and Kelvin Lee, Kin Long and Charnley, Steven B. and Cordiner, Martin A. and Herbst, Eric and Kalenskii, Sergei and Momjian, Emmanuel and Willis, Eric R. and Xue, Ci and Remijan, Anthony J. and McCarthy, Michael C.},
  year = {2020},
  month = sep,
  journal = {The Astrophysical Journal Letters},
  volume = {900},
  number = {1},
  pages = {L10},
  publisher = {The American Astronomical Society},
  issn = {2041-8205},
  doi = {10.3847/2041-8213/aba632},
  urldate = {2025-05-23},
}

@misc{molsim,
  title = {Molsim},
  author = {McGuire, Brett A. and Xue, Ci and Lee, Kin Long Kelvin and {El-Abd}, Samer and Loomis, Ryan A.},
  year = 2024,
  month = jul,
  doi = {10.5281/zenodo.12697227},
  urldate = {2026-01-13},
  howpublished = {Zenodo}
}

@article{stacking,
  title = {An Investigation of Spectral Line Stacking Techniques and Application to the Detection of {{HC11N}}},
  author = {Loomis, Ryan A. and Burkhardt, Andrew M. and Shingledecker, Christopher N. and Charnley, Steven B. and Cordiner, Martin A. and Herbst, Eric and Kalenskii, Sergei and Lee, Kin Long Kelvin and Willis, Eric R. and Xue, Ci and Remijan, Anthony J. and McCarthy, Michael C. and McGuire, Brett A.},
  year = 2021,
  month = feb,
  journal = {Nature Astronomy},
  volume = {5},
  number = {2},
  pages = {188--196},
  publisher = {Nature Publishing Group},
  issn = {2397-3366},
  doi = {10.1038/s41550-020-01261-4},
  urldate = {2026-01-15},
  copyright = {2021 The Author(s), under exclusive licence to Springer Nature Limited},
  langid = {english}
}

@article{gasphasekinetics,
  title = {Gas-Phase {{Chemistry}} in the {{Interstellar Medium}}: {{The Role}} of {{Laboratory Astrochemistry}}},
  shorttitle = {Gas-Phase {{Chemistry}} in the {{Interstellar Medium}}},
  author = {Puzzarini, Cristina},
  year = 2022,
  month = feb,
  journal = {Frontiers in Astronomy and Space Sciences},
  volume = {8},
  publisher = {Frontiers},
  issn = {2296-987X},
  doi = {10.3389/fspas.2021.811342},
  urldate = {2025-07-25},
  langid = {english}
}

@article{H2C3O1,
  title = {{{INVESTIGATING THE MINIMUM ENERGY PRINCIPLE IN SEARCHES FOR NEW MOLECULAR SPECIES}}---{{THE CASE OF H2C3O ISOMERS}}},
  author = {Loomis, Ryan A. and McGuire, Brett A. and Shingledecker, Christopher and Johnson, Chelen H. and Blair, Samantha and Robertson, Amy and Remijan, Anthony J.},
  year = 2015,
  month = jan,
  journal = {The Astrophysical Journal},
  volume = {799},
  number = {1},
  pages = {34},
  publisher = {The American Astronomical Society},
  issn = {0004-637X},
  doi = {10.1088/0004-637X/799/1/34},
  urldate = {2026-01-21},
  langid = {english}
}

@article{H2C3O2,
  title = {The Interstellar Chemistry of {{H}}{\textsubscript{2}}{{C}}{\textsubscript{3}}{{O}} Isomers},
  author = {Loison, Jean-Christophe and Ag{\'u}ndez, Marcelino and Marcelino, N{\'u}ria and Wakelam, Valentine and Hickson, Kevin M. and Cernicharo, Jos{\'e} and Gerin, Maryvonne and Roueff, Evelyne and Gu{\'e}lin, Michel},
  year = 2016,
  month = mar,
  journal = {Monthly Notices of the Royal Astronomical Society},
  volume = {456},
  number = {4},
  pages = {4101--4110},
  publisher = {Oxford University Press (OUP)},
  issn = {0035-8711, 1365-2966},
  doi = {10.1093/mnras/stv2866},
  urldate = {2025-07-20},
  langid = {english}
}

@article{H2C3O3,
  title = {Discovery of Linear Propadienone: {{Study}} of the Chemistry of Linear and Cyclic {{H}}{\textsubscript{2}} {{C}}{\textsubscript{3}} {{O}} and {{H}}{\textsubscript{2}} {{C}}{\textsubscript{3}} {{S}} in {{TMC-1}}},
  shorttitle = {Discovery of Linear Propadienone},
  author = {Esplugues, G. and Loison, J. C. and Ag{\'u}ndez, M. and Molpeceres, G. and Marcelino, N. and Tercero, B. and Cernicharo, J.},
  year = 2026,
  month = jan,
  journal = {Astronomy \& Astrophysics},
  volume = {705},
  pages = {A206},
  issn = {0004-6361, 1432-0746},
  doi = {10.1051/0004-6361/202557355},
  urldate = {2026-03-27},
  copyright = {https://creativecommons.org/licenses/by/4.0},
  langid = {english}
}

@article{HACA1,
  title = {Hydrogen {{Abstraction}}/{{Acetylene Addition Revealed}}},
  author = {Parker, Dorian S. N. and Kaiser, Ralf I. and Troy, Tyler P. and Ahmed, Musahid},
  year = 2014,
  journal = {Angewandte Chemie International Edition},
  volume = {53},
  number = {30},
  pages = {7740--7744},
  issn = {1521-3773},
  doi = {10.1002/anie.201404537},
  urldate = {2026-01-21},
  copyright = {\copyright{} 2014 WILEY-VCH Verlag GmbH \& Co. KGaA, Weinheim},
  langid = {english}
}

@article{C3addition1,
  title = {Formation {{Mechanisms}} of {{Naphthalene}} and {{Indene}}: {{From}} the {{Interstellar Medium}} to {{Combustion Flames}}},
  shorttitle = {Formation {{Mechanisms}} of {{Naphthalene}} and {{Indene}}},
  author = {Mebel, Alexander M. and Landera, Alexander and Kaiser, Ralf I.},
  year = 2017,
  month = feb,
  journal = {The Journal of Physical Chemistry A},
  volume = {121},
  number = {5},
  pages = {901--926},
  publisher = {American Chemical Society},
  issn = {1089-5639},
  doi = {10.1021/acs.jpca.6b09735},
  urldate = {2026-01-21}
}

@article{HAVA1,
  title = {Low Temperature Formation of Naphthalene and Its Role in the Synthesis of {{PAHs}} ({{Polycyclic Aromatic Hydrocarbons}}) in the Interstellar Medium},
  author = {Parker, Dorian S. N. and Zhang, Fangtong and Kim, Y. Seol and Kaiser, Ralf I. and Landera, Alexander and Kislov, Vadim V. and Mebel, Alexander M. and Tielens, A. G. G. M.},
  year = {2012},
  journal = {Proceedings of the National Academy of Sciences},
  volume = {109},
  number = {1},
  pages = {53--58},
  publisher = {Proceedings of the National Academy of Sciences},
  doi = {10.1073/pnas.1113827108}
}

@article{PAC,
  title = {Synthesis of {{Polycyclic Aromatic Hydrocarbons}} by {{Phenyl Addition}}--{{Dehydrocyclization}}: {{The Third Way}}},
  shorttitle = {Synthesis of {{Polycyclic Aromatic Hydrocarbons}} by {{Phenyl Addition}}--{{Dehydrocyclization}}},
  author = {Zhao, Long and Prendergast, Matthew B. and Kaiser, Ralf I. and Xu, Bo and Ablikim, Utuq and Ahmed, Musahid and Sun, Bing-Jian and Chen, Yue-Lin and Chang, Agnes H. H. and Mohamed, Rana K. and Fischer, Felix R.},
  year = 2019,
  journal = {Angewandte Chemie International Edition},
  volume = {58},
  number = {48},
  pages = {17442--17450},
  issn = {1521-3773},
  doi = {10.1002/anie.201909876},
  urldate = {2026-01-21},
  copyright = {\copyright{} 2019 Wiley-VCH Verlag GmbH \& Co. KGaA, Weinheim},
  langid = {english}
}

@article{review,
  title = {An {{Aromatic Universe}}--{{A Physical Chemistry Perspective}}},
  author = {Kaiser, Ralf I. and Hansen, Nils},
  year = 2021,
  month = may,
  journal = {The Journal of Physical Chemistry A},
  volume = {125},
  number = {18},
  pages = {3826--3840},
  issn = {1089-5639, 1520-5215},
  doi = {10.1021/acs.jpca.1c00606},
  urldate = {2026-01-22},
  copyright = {https://doi.org/10.15223/policy-029},
  langid = {english}
}

@article{nath2phen,
  title = {Low-Temperature Formation of Polycyclic Aromatic Hydrocarbons in {{Titan}}'s Atmosphere},
  author = {Zhao, Long and Kaiser, Ralf I. and Xu, Bo and Ablikim, Utuq and Ahmed, Musahid and Evseev, Mikhail M. and Bashkirov, Eugene K. and Azyazov, Valeriy N. and Mebel, Alexander M.},
  year = 2018,
  month = dec,
  journal = {Nature Astronomy},
  volume = {2},
  number = {12},
  pages = {973--979},
  publisher = {Nature Publishing Group},
  issn = {2397-3366},
  doi = {10.1038/s41550-018-0585-y},
  urldate = {2026-01-22},
  copyright = {2018 The Author(s), under exclusive licence to Springer Nature Limited},
  langid = {english}
}

@article{phen2chry,
  title = {A {{Unified Mechanism}} on the {{Formation}} of {{Acenes}}, {{Helicenes}}, and {{Phenacenes}} in the {{Gas Phase}}},
  author = {Zhao, Long and Kaiser, Ralf I. and Xu, Bo and Ablikim, Utuq and Ahmed, Musahid and Evseev, Mikhail M. and Bashkirov, Eugene K. and Azyazov, Valeriy N. and Mebel, Alexander M.},
  year = 2020,
  journal = {Angewandte Chemie International Edition},
  volume = {59},
  number = {10},
  pages = {4051--4058},
  issn = {1521-3773},
  doi = {10.1002/anie.201913037},
  urldate = {2026-01-22},
  copyright = {\copyright{} 2019 Wiley-VCH Verlag GmbH \& Co. KGaA, Weinheim},
  langid = {english}
}

@article{CH,
  title = {Low-Temperature Gas-Phase Formation of Indene in the Interstellar Medium},
  author = {Doddipatla, Srinivas and Galimova, Galiya R. and Wei, Hongji and Thomas, Aaron M. and He, Chao and Yang, Zhenghai and Morozov, Alexander N. and Shingledecker, Christopher N. and Mebel, Alexander M. and Kaiser, Ralf I.},
  year = 2021,
  month = jan,
  journal = {Science Advances},
  volume = {7},
  number = {1},
  pages = {eabd4044},
  publisher = {American Association for the Advancement of Science},
  doi = {10.1126/sciadv.abd4044},
  urldate = {2026-01-27}
}

@article{C3,
  title = {Reaction of the {{C3}} Carbon Cluster with Benzene},
  author = {Szczepanski, Jan and Wang, Haiyan and Vala, Martin},
  year = 2004,
  month = aug,
  journal = {Chemical Physics},
  volume = {303},
  number = {1},
  pages = {165--177},
  issn = {0301-0104},
  doi = {10.1016/j.chemphys.2004.04.028},
  urldate = {2026-01-27}
}

@article{C2,
  title = {A Combined Crossed Molecular Beam and Ab Initio Investigation of {{C2}} and {{C3}} Elementary Reactions with Unsaturated Hydrocarbons---Pathways to Hydrogen Deficient Hydrocarbon Radicals in Combustion Flames},
  author = {Kaiser, Ralf I. and Le, Trung N. and Nguyen, Thanh L. and Mebel, Alexander M. and Balucani, Nadia and Lee, Yuan T. and Stahl, Frank and Schleyer, Paul v R. and Iii, Henry F. Schaefer},
  year = 2002,
  month = nov,
  journal = {Faraday Discussions},
  volume = {119},
  number = {0},
  pages = {51--66},
  publisher = {The Royal Society of Chemistry},
  issn = {1364-5498},
  doi = {10.1039/B101967H},
  urldate = {2026-01-27},
  langid = {english}
}

@article{C2HtoBenzene,
  title = {Formation of Benzene in the Interstellar Medium},
  author = {Jones, Brant M. and Zhang, Fangtong and Kaiser, Ralf I. and Jamal, Adeel and Mebel, Alexander M. and Cordiner, Martin A. and Charnley, Steven B.},
  year = 2011,
  month = jan,
  journal = {Proceedings of the National Academy of Sciences},
  volume = {108},
  number = {2},
  pages = {452--457},
  publisher = {Proceedings of the National Academy of Sciences},
  doi = {10.1073/pnas.1012468108},
  urldate = {2026-01-27}
}

@article{indeneHAVA,
  title = {Reactivity of the {{Indenyl Radical}} ({{C9H7}}) with {{Acetylene}} ({{C2H2}}) and {{Vinylacetylene}} ({{C4H4}})},
  author = {Zhao, Long and Prendergast, Matthew B. and Kaiser, Ralf I. and Xu, Bo and Lu, Wenchao and Ablikim, Utuq and Ahmed, Musahid and Oleinikov, Artem D. and Azyazov, Valeriy N. and Mebel, Alexander M. and Howlader, A. Hasan and Wnuk, Stanislaw F.},
  year = 2019,
  journal = {ChemPhysChem},
  volume = {20},
  number = {11},
  pages = {1437--1447},
  issn = {1439-7641},
  doi = {10.1002/cphc.201900052},
  urldate = {2026-01-27},
  copyright = {\copyright{} 2019 Wiley-VCH Verlag GmbH \& Co. KGaA, Weinheim},
  langid = {english}
}

@article{CN+bezene,
  title = {Benzonitrile as a {{Proxy}} for {{Benzene}} in the {{Cold ISM}}: {{Low-temperature Rate Coefficients}} for {{CN}} + {{C6H6}}},
  shorttitle = {Benzonitrile as a {{Proxy}} for {{Benzene}} in the {{Cold ISM}}},
  author = {Cooke, Ilsa R. and Gupta, Divita and Messinger, Joseph P. and Sims, Ian R.},
  year = 2020,
  month = mar,
  journal = {The Astrophysical Journal Letters},
  volume = {891},
  number = {2},
  pages = {L41},
  publisher = {The American Astronomical Society},
  issn = {2041-8205},
  doi = {10.3847/2041-8213/ab7a9c},
  urldate = {2026-01-27},
  langid = {english}
}

@article{cyclopentindene,
  title = {Discovery of 1 {{{\emph{H}}}} -Cyclopent[ {\emph{Cd}} ]Indene ( {\emph{c}} -{{C}}{\textsubscript{11}} {{H}}{\textsubscript{8}} ) in {{TMC-1}} with the {{QUIJOTE}} Line Survey: {{A}} New Three-Ringed Polycyclic Aromatic Hydrocarbon},
  shorttitle = {Discovery of 1 {{{\emph{H}}}} -Cyclopent[ {\emph{Cd}} ]Indene ( {\emph{c}} -{{C}}{\textsubscript{11}} {{H}}{\textsubscript{8}} ) in {{TMC-1}} with the {{QUIJOTE}} Line Survey},
  author = {Fuentetaja, R. and Cabezas, C. and Ag{\'u}ndez, M. and Tercero, B. and Marcelino, N. and De Vicente, P. and Cernicharo, J.},
  year = 2026,
  month = feb,
  journal = {Astronomy \& Astrophysics},
  volume = {706},
  pages = {L10},
  issn = {0004-6361, 1432-0746},
  doi = {10.1051/0004-6361/202558398},
  urldate = {2026-07-28},
  copyright = {https://creativecommons.org/licenses/by/4.0},
  langid = {english}
}

@article{cyanoacenaphthylene2,
  title = {Discovery of Two New Isomers of Cyanoacenaphthylene ({{C}}{\textsubscript{12}} {{H}}{\textsubscript{7}} {{CN}}) in the {{Taurus}} Molecular Cloud 1 with the {{QUIJOTE}} Line Survey},
  author = {Cernicharo, J. and Tercero, B. and Marcelino, N. and {L{\'o}pez-P{\'e}rez}, J. A. and Gallego, J. D. and Tercero, F. and Esplugues, G. and Cabezas, C. and Ag{\'u}ndez, M. and Limeres, C. and Steber, A. L. and P{\'e}rez, D. and P{\'e}rez, C. and Lesarri, A. and De Vicente, P.},
  year = 2026,
  month = jan,
  journal = {Astronomy \& Astrophysics},
  volume = {705},
  pages = {L7},
  issn = {0004-6361, 1432-0746},
  doi = {10.1051/0004-6361/202557893},
  urldate = {2026-02-05},
  copyright = {https://creativecommons.org/licenses/by/4.0},
  langid = {english}
}

@article{orionbarJWSTPAH,
  title = {{{PDRs4All}}-{{XIV}}. Probing {{CH}} out-of-plane bending modes of {{PAH}} molecules in the {{Orion Bar}} with {{JWST}}},
  author = {Khan, Yasir and Bern{\'e}, Olivier and Croiset, Bastien A. and Chastenet, Juliette and Joblin, Christine and Peeters, Els and Sidhu, Amanpreet and van De Putte, Dave and Abergel, Alain and Arab, Hiba and Bally, John and Boersma, Christiaan and Canin, Gregory and Cami, Jan and Cazaux, St{\'e}phane and Habart, Edith and Lagache, Guilaine and Maragkoudakis, Alexandros and Pilleri, Paolo and Tielens, A. G. G. M. and Van de Putte, David and Zhen, J.},
  year = 2025,
  journal = {Astronomy \& Astrophysics},
  volume = {699},
  pages = {A133},
  doi = {10.1051/0004-6361/202554096},
  langid = {english}
}

@article{Hony2001,
  title = {The CH out-of-plane bending modes of polycyclic aromatic hydrocarbons and astrophysical implications},
  author = {Hony, S. and Van Kerckhoven, C. and Peeters, E. and Hudgins, D. M. and Allamandola, L. J. and Tielens, A. G. G. M.},
  year = 2001,
  journal = {Astronomy \& Astrophysics},
  volume = {370},
  pages = {1030--1043},
  doi = {10.1051/0004-6361:20010242},
  langid = {english}
}

@article{Peeters2002,
  title = {The rich 6 to 9 {$\mu$m} spectrum of interstellar PAHs},
  author = {Peeters, E. and Hony, S. and Van Kerckhoven, C. and Tielens, A. G. G. M. and Allamandola, L. J. and Hudgins, D. M. and Bauschlicher, C. W.},
  year = 2002,
  journal = {Astronomy \& Astrophysics},
  volume = {390},
  pages = {1089--1113},
  doi = {10.1051/0004-6361:20020773},
  langid = {english}
}

@article{Ricca2019,
  title = {Toward the Identification of the Carriers of the Second Aromatic Infrared Band Spectral Class},
  author = {Ricca, A. and Bauschlicher, C. W. and Boersma, C. and Peeters, E. and Tielens, A. G. G. M. and Allamandola, L. J.},
  year = 2019,
  journal = {The Astrophysical Journal},
  volume = {882},
  pages = {56},
  doi = {10.3847/1538-4357/ab3124},
  langid = {english}
}

@article{Chown2024,
  title = {The changing shape of the 11.2 {$\mu$m} emission band in the Orion Bar},
  author = {Chown, R. and Peeters, E. and Shannon, M. J. and Doney, K. D. and Tielens, A. G. G. M. and Bern{\'e}, O. and Cami, J. and Croiset, B. A. and Maragkoudakis, A. and Pilleri, P.},
  year = 2024,
  journal = {Astronomy \& Astrophysics},
  volume = {685},
  pages = {A75},
  doi = {10.1051/0004-6361/202346662},
  langid = {english}
}

@article{vanDiedenhoven2004,
  title = {The Profiles of the 3.3 {$\mu$m} and 11.2 {$\mu$m} {{PAH}} Features},
  author = {van Diedenhoven, Bastiaan and Peeters, Els and Van Kerckhoven, Christiaan and Hony, S. and Hudgins, Douglas M. and Allamandola, Louis J. and Tielens, A. G. G. M.},
  year = 2004,
  journal = {The Astrophysical Journal},
  volume = {611},
  pages = {928--939},
  doi = {10.1086/422404},
  langid = {english}
}

@article{TriphenyleneDIBs,
  title = {Laboratory Spectroscopy and Astronomical Significance of the Fully-Benzenoid {{PAH}} Triphenylene and Its Cation},
  author = {Kofman, V. and Sarre, P. J. and Hibbins, R. E. and {ten Kate}, I. L. and Linnartz, H.},
  year = 2017,
  month = jun,
  journal = {Molecular Astrophysics},
  volume = {7},
  pages = {19--26},
  issn = {2405-6758},
  doi = {10.1016/j.molap.2017.04.002},
  urldate = {2025-12-19}
}

@article{ORCA:2025,
  author = {Neese, Frank},
  title = {Software Update: The ORCA Program System, Version 6.0},
  journal = {WIREs Computational Molecular Science},
  year = {2025},
  month = mar,
  volume = {15},
  number = {2},
  publisher = {Wiley},
  issn = {1759-0876, 1759-0884},
  doi = {10.1002/wcms.70019},
  urldate = {2025-07-16}
}

@article{SHARK:2023,
	title = {The {SHARK} integral generation and digestion system},
	volume = {44},
	copyright = {漏 2022 The Author. Journal of Computational Chemistry published by Wiley Periodicals LLC.},
	issn = {1096-987X},
	url = {https://onlinelibrary.wiley.com/doi/abs/10.1002/jcc.26942},
	doi = {10.1002/jcc.26942},
	language = {en},
	number = {3},
	urldate = {2026-01-19},
	journal = {Journal of Computational Chemistry},
	author = {Neese, Frank},
	year = {2023},
	note = {\_eprint: https://onlinelibrary.wiley.com/doi/pdf/10.1002/jcc.26942},
	pages = {381--396},
}

@article{ORCA:2012,
	title = {The {ORCA} program system},
	volume = {2},
	copyright = {Copyright 漏 2011 John Wiley \& Sons, Ltd.},
	issn = {1759-0884},
	url = {https://onlinelibrary.wiley.com/doi/abs/10.1002/wcms.81},
	doi = {10.1002/wcms.81},
	language = {en},
	number = {1},
	urldate = {2026-01-19},
	journal = {WIREs Computational Molecular Science},
	author = {Neese, Frank},
	year = {2012},
	note = {\_eprint: https://wires.onlinelibrary.wiley.com/doi/pdf/10.1002/wcms.81},
	pages = {73--78},
}

@article{UHF:2000,
	title = {Approximate second-order {SCF} convergence for spin unrestricted wavefunctions},
	volume = {325},
	issn = {0009-2614},
	url = {https://www.sciencedirect.com/science/article/pii/S000926140000662X},
	doi = {10.1016/S0009-2614(00)00662-X},
	number = {1},
	urldate = {2026-01-19},
	journal = {Chemical Physics Letters},
	author = {Neese, Frank},
	month = jul,
	year = {2000},
	pages = {93--98},
}

@article{Perdew:1986:8822,
    title = {Density-functional approximation for the correlation energy of the inhomogeneous electron gas},
    volume = {33},
    url = {https://link.aps.org/doi/10.1103/PhysRevB.33.8822},
    doi = {10.1103/PhysRevB.33.8822},
    number = {12},
    urldate = {2025-03-06},
    journal = {Physical Review B},
    author = {Perdew, John P.},
    month = jun,
    year = {1986},
    note = {Publisher: American Physical Society},
    pages = {8822--8824},
}

@article{C3H6O2,
  title = {Expanding the {{C3H6O2}} Isomeric Interstellar Inventory: {{Discovery}} of Lactaldehyde and Methoxyacetaldehyde in {{G}}+0.693-0.027},
  shorttitle = {Expanding the {{C3H6O2}} Isomeric Interstellar Inventory},
  author = {{Sanz-Novo}, M. and Rivilla, V. M. and {Jim{\'e}nez-Serra}, I. and Colzi, L. and Zeng, S. and Meg{\'i}as, A. and Andr{\'e}s, D. San and {L{\'o}pez-Gallifa}, {\'A} and {Mart{\'i}nez-Henares}, A. and Fried, Z. T. P. and McGuire, B. A. and Mart{\'i}n, S. and {Requena-Torres}, M. A. and Tercero, B. and de Vicente, P. and Kolesnikov{\'a}, L. and Alonso, E. R. and Cocinero, E. J. and Guillemin, J. C. and Kleiner, I.},
  year = 2026,
  month = feb,
  journal = {Astronomy \& Astrophysics},
  volume = {706},
  pages = {A316},
  publisher = {EDP Sciences},
  issn = {0004-6361, 1432-0746},
  doi = {10.1051/0004-6361/202558316},
  urldate = {2026-03-18},
  copyright = {\copyright{} The Authors 2026},
  langid = {english}
}

@article{electronicstability,
  title = {Structure-Dependent {{Photostability}} of {{Polycyclic Aromatic Hydrocarbon Cations}}: {{Laboratory Studies}} and {{Astrophysical Implications}}},
  shorttitle = {Structure-Dependent {{Photostability}} of {{Polycyclic Aromatic Hydrocarbon Cations}}},
  author = {Jochims, H. W. and Baumg{\"a}rtel, H. and Leach, S.},
  year = 1999,
  month = feb,
  journal = {The Astrophysical Journal},
  volume = {512},
  number = {1},
  pages = {500},
  publisher = {IOP Publishing},
  issn = {0004-637X},
  doi = {10.1086/306752},
  urldate = {2026-03-18},
  langid = {english}
}

@article{Nath2Acey,
  title = {Hydrogen {{Abstraction Acetylene Addition}} and {{Diels}}-{{Alder Mechanisms}} of {{PAH Formation}}:\, {{A Detailed Study Using First Principles Calculations}}},
  shorttitle = {Hydrogen {{Abstraction Acetylene Addition}} and {{Diels}}-{{Alder Mechanisms}} of {{PAH Formation}}},
  author = {Kislov, V. V. and Islamova, N. I. and Kolker, A. M. and Lin, S. H. and Mebel, A. M.},
  year = 2005,
  month = sep,
  journal = {Journal of Chemical Theory and Computation},
  volume = {1},
  number = {5},
  pages = {908--924},
  publisher = {American Chemical Society},
  issn = {1549-9618},
  doi = {10.1021/ct0500491},
  urldate = {2026-03-18}
}

@article{Nath2Phan,
  title = {Gas Phase Formation of Phenalene via 10{$\pi$}-Aromatic, Resonantly Stabilized Free Radical Intermediates},
  author = {Zhao, Long and I.~Kaiser, Ralf and Lu, Wenchao and Ahmed, Musahid and D.~Oleinikov, Artem and N.~Azyazov, Valeriy and M.~Mebel, Alexander and Hasan~Howlader, A. and F.~Wnuk, Stanislaw},
  year = 2020,
  journal = {Physical Chemistry Chemical Physics},
  volume = {22},
  number = {27},
  pages = {15381--15388},
  publisher = {Royal Society of Chemistry},
  doi = {10.1039/D0CP02216K},
  urldate = {2026-03-18},
  langid = {english}
}

@article{Langhoff1996,
  author = {Langhoff, Stephen R.},
  title = {Theoretical Infrared Spectra for Polycyclic Aromatic Hydrocarbon Neutrals, Cations, and Anions},
  journal = {The Journal of Physical Chemistry},
  year = 1996,
  volume = {100},
  number = {8},
  pages = {2819--2841},
  doi = {10.1021/jp952074g},
}

@article{triphenylenelesscompact,
  title = {Polycyclic Aromatic Hydrocarbon Molecules and the {{2175\AA}} Interstellar Extinction Bump},
  author = {Lin, Qi and Yang, X J and Li, Aigen},
  year = 2023,
  month = aug,
  journal = {Monthly Notices of the Royal Astronomical Society},
  volume = {525},
  number = {2},
  pages = {2380--2387},
  issn = {0035-8711, 1365-2966},
  doi = {10.1093/mnras/stad2405},
  urldate = {2026-03-19},
  copyright = {https://academic.oup.com/journals/pages/open\_access/funder\_policies/chorus/standard\_publication\_model},
  langid = {english}
}

@article{ethynyladdition,
  title = {Photoinduced {{Mechanism}} of {{Formation}} and {{Growth}} of {{Polycyclic Aromatic Hydrocarbons}} in {{Low-Temperature Environments}} via {{Successive Ethynyl Radical Additions}}},
  author = {Mebel, Alexander M. and Kislov, Vadim V. and Kaiser, Ralf I.},
  year = 2008,
  month = oct,
  journal = {Journal of the American Chemical Society},
  volume = {130},
  number = {41},
  pages = {13618--13629},
  publisher = {American Chemical Society},
  issn = {0002-7863},
  doi = {10.1021/ja804198a},
  urldate = {2026-03-19}
}

@article{ethynylcyclopentadiene,
  title = {Discovery of Two Isomers of Ethynyl Cyclopentadiene in {{TMC-1}}: {{Abundances}} of {{CCH}} and {{CN}} Derivatives of Hydrocarbon Cycles},
  shorttitle = {Discovery of Two Isomers of Ethynyl Cyclopentadiene in {{TMC-1}}},
  author = {Cernicharo, J. and Ag{\'u}ndez, M. and Kaiser, R. I. and Cabezas, C. and Tercero, B. and Marcelino, N. and Pardo, J. R. and De Vicente, P.},
  year = 2021,
  month = nov,
  journal = {Astronomy \& Astrophysics},
  volume = {655},
  pages = {L1},
  issn = {0004-6361, 1432-0746},
  doi = {10.1051/0004-6361/202142226},
  urldate = {2025-04-17},
  copyright = {https://www.edpsciences.org/en/authors/copyright-and-licensing},
  langid = {english}
}

@article{ethynylbenzene,
  title = {Detection of Ethynylbenzene in {{TMC-1}} and the Interstellar Search for 1,2-Diethynylbenzene},
  author = {Loru, Donatella and Cabezas, Carlos and Cernicharo, Jos{\'e} and Schnell, Melanie and Steber, Amanda L.},
  year = 2023,
  month = sep,
  journal = {Astronomy \& Astrophysics},
  volume = {677},
  pages = {A166},
  publisher = {EDP Sciences},
  issn = {0004-6361, 1432-0746},
  doi = {10.1051/0004-6361/202347023},
  urldate = {2026-03-19},
  copyright = {\copyright{} The Authors 2023},
  langid = {english}
}

@article{CCHCNratio,
  title = {A {{Study}} of the {{Physics}} and {{Chemistry}} of {{TMC-1}}},
  author = {Pratap, P. and Dickens, J. E. and Snell, R. L. and Miralles, M. P. and Bergin, E. A. and Irvine, W. M. and Schloerb, F. P.},
  year = 1997,
  month = sep,
  journal = {The Astrophysical Journal},
  volume = {486},
  number = {2},
  pages = {862},
  publisher = {IOP Publishing},
  issn = {0004-637X},
  doi = {10.1086/304553},
  urldate = {2026-03-19},
  langid = {english}
}

@article{l-C3H,
  title = {Astronomical Identification of the {{C3 H}} Radical.},
  author = {Thaddeus, P. and Gottlieb, C. A. and Hjalmarson, A. and Johansson, L. E. B. and Irvine, W. M. and Friberg, P. and Linke, R. A.},
  year = 1985,
  month = jul,
  journal = {The Astrophysical Journal},
  volume = {294},
  pages = {L49-L53},
  publisher = {IOP},
  issn = {0004-637X},
  doi = {10.1086/184507},
  urldate = {2026-03-19}
}

@article{c-C3H,
  title = {Observations of the Cyclic {{C3H}} Radical in the Interstellar Medium.},
  author = {Mangum, J. G. and Wootten, A.},
  year = 1990,
  month = nov,
  journal = {Astronomy and Astrophysics},
  volume = {239},
  pages = {319},
  publisher = {EDP},
  issn = {0004-6361},
  urldate = {2026-03-19}
}

@article{C3H2,
  title = {Molecular {{Carbon Chains}} and {{Rings}} in {{TMC-1}}},
  author = {Foss{\'e}, David and Cernicharo, Jos{\'e} and Gerin, Maryvonne and Cox, Pierre},
  year = 2001,
  month = may,
  journal = {The Astrophysical Journal},
  volume = {552},
  number = {1},
  pages = {168},
  publisher = {IOP Publishing},
  issn = {0004-637X},
  doi = {10.1086/320471},
  urldate = {2026-03-19},
  langid = {english}
}

@article{C3H2-2,
  title = {The {{Physics}} and {{Chemistry}} of {{Small Translucent Molecular Clouds}}. {{IX}}. {{Acetylenic Chemistry}}},
  author = {Turner, B. E. and Lee, Ho-Hsin and Herbst, Eric},
  year = 1998,
  month = mar,
  journal = {The Astrophysical Journal Supplement Series},
  volume = {115},
  number = {1},
  pages = {91},
  publisher = {IOP Publishing},
  issn = {0067-0049},
  doi = {10.1086/313077},
  urldate = {2026-03-19},
  langid = {english}
}

@article{C3H3,
  title = {Discovery of the Propargyl Radical ({{CH}}{\textsubscript{2}} {{CCH}}) in {{TMC-1}}: {{One}} of the Most Abundant Radicals Ever Found and a Key Species for Cyclization to Benzene in Cold Dark Clouds},
  shorttitle = {Discovery of the Propargyl Radical ({{CH}}{\textsubscript{2}} {{CCH}}) in {{TMC-1}}},
  author = {Ag{\'u}ndez, M. and Cabezas, C. and Tercero, B. and Marcelino, N. and Gallego, J. D. and De Vicente, P. and Cernicharo, J.},
  year = 2021,
  month = mar,
  journal = {Astronomy \& Astrophysics},
  volume = {647},
  pages = {L10},
  issn = {0004-6361, 1432-0746},
  doi = {10.1051/0004-6361/202140553},
  urldate = {2026-03-19},
  copyright = {https://www.edpsciences.org/en/authors/copyright-and-licensing},
  langid = {english}
}

@article{Hloss,
  title = {Photoinduced Polycyclic Aromatic Hydrocarbon Dehydrogenation {{The}} Competition between {{H-}} and {{H}}{\textsubscript{2}} -Loss},
  author = {Castellanos, P. and Candian, A. and Zhen, J. and Linnartz, H. and Tielens, A. G. G. M.},
  year = 2018,
  month = aug,
  journal = {Astronomy \& Astrophysics},
  volume = {616},
  pages = {A166},
  issn = {0004-6361, 1432-0746},
  doi = {10.1051/0004-6361/201833220},
  urldate = {2026-03-19},
  copyright = {https://www.edpsciences.org/en/authors/copyright-and-licensing},
  langid = {english}
}

@article{D3,
  title = {A Consistent and Accurate Ab Initio Parametrization of Density Functional Dispersion Correction ({{DFT-D}}) for the 94 Elements {{H-Pu}}},
  author = {Grimme, Stefan and Antony, Jens and Ehrlich, Stephan and Krieg, Helge},
  year = {2010},
  month = apr,
  journal = {The Journal of Chemical Physics},
  volume = {132},
  number = {15},
  pages = {154104},
  issn = {0021-9606},
  doi = {10.1063/1.3382344},
  urldate = {2025-08-12},
}

@article{def2TZVPP1,
  title = {Accurate {{Coulomb-fitting}} Basis Sets for {{H}} to {{Rn}}},
  author = {Weigend, Florian},
  year = {2006},
  month = feb,
  journal = {Physical Chemistry Chemical Physics},
  volume = {8},
  number = {9},
  pages = {1057--1065},
  publisher = {The Royal Society of Chemistry},
  issn = {1463-9084},
  doi = {10.1039/B515623H},
  urldate = {2025-08-12},
  langid = {english},
}

@article{def2TZVPP2,
  title = {Balanced Basis Sets of Split Valence, Triple Zeta Valence and Quadruple Zeta Valence Quality for {{H}} to {{Rn}}: {{Design}} and Assessment of Accuracy},
  shorttitle = {Balanced Basis Sets of Split Valence, Triple Zeta Valence and Quadruple Zeta Valence Quality for {{H}} to {{Rn}}},
  author = {Weigend, Florian and Ahlrichs, Reinhart},
  year = {2005},
  month = aug,
  journal = {Physical Chemistry Chemical Physics},
  volume = {7},
  number = {18},
  pages = {3297--3305},
  publisher = {The Royal Society of Chemistry},
  issn = {1463-9084},
  doi = {10.1039/B508541A},
  urldate = {2025-08-12},
  langid = {english},
}

@article{1-CNTNMR,
  title = {Cu-{{Catalyzed Aromatic Metamorphosis}} of 3-{{Aminoindazoles}}},
  author = {Zhou, Yao and Lin, Lu and Wang, Ya and Zhu, Jun and Song, Qiuling},
  year = 2019,
  month = sep,
  journal = {Organic Letters},
  volume = {21},
  number = {18},
  pages = {7630--7634},
  issn = {1523-7060, 1523-7052},
  doi = {10.1021/acs.orglett.9b02933},
  urldate = {2026-04-30},
  copyright = {https://doi.org/10.15223/policy-029},
  langid = {english}
}

@article{orca2020,
  title = {The {{ORCA}} Quantum Chemistry Program Package},
  author = {Neese, Frank and Wennmohs, Frank and Becker, Ute and Riplinger, Christoph},
  year = {2020},
  month = jun,
  journal = {The Journal of Chemical Physics},
  volume = {152},
  number = {22},
  pages = {224108},
  publisher = {AIP Publishing},
  issn = {0021-9606, 1089-7690},
  doi = {10.1063/5.0004608},
  urldate = {2025-07-16},
}

@article{cyanofluoranthene,
  title = {Rotational {{Characterization}} of {{Four-Ring Polycyclic Aromatic Hydrocarbons}}: {{Toward}} the {{Detection}} of {{Fluoranthene}} and {{Cyanofluoranthene}} in {{Space}}},
  shorttitle = {Rotational {{Characterization}} of {{Four-Ring Polycyclic Aromatic Hydrocarbons}}},
  author = {{Villar-Castro}, Daniel and Cabezas, Carlos and Steber, Amanda L. and Mor{\'a}n, Jos{\'e} R. and {de la Fuente}, Selene and Hussain, Farha S. and P{\'e}rez, Dolores and Lesarri, Alberto and Cernicharo, Jos{\'e} and P{\'e}rez, Crist{\'o}bal and Pe{\~n}a, Isabel},
  year = 2026,
  month = apr,
  journal = {The Journal of Physical Chemistry Letters},
  volume = {17},
  number = {17},
  pages = {5012--5020},
  publisher = {American Chemical Society},
  doi = {10.1021/acs.jpclett.6c00413},
  urldate = {2026-05-11}
}

@article{cyanobiphenylene,
  title = {Cyano-{{Polycyclic Aromatic Hydrocarbon Interstellar Candidates}}: {{Laboratory Identification}}, {{Equilibrium Structure}} and {{Astronomical Search}} of {{Cyanobiphenylene}}},
  shorttitle = {Cyano-{{Polycyclic Aromatic Hydrocarbon Interstellar Candidates}}},
  author = {Cabezas, Carlos and Janeiro, Jes{\'u}s and P{\'e}rez, Dolores and Li, Wenqin and Ag{\'u}ndez, Marcelino and Steber, Amanda L. and Guiti{\'a}n, Enrique and Demaison, Jean and P{\'e}rez, Crist{\'o}bal and Cernicharo, Jos{\'e} and Lesarri, Alberto},
  year = 2024,
  month = jul,
  journal = {The Journal of Physical Chemistry Letters},
  volume = {15},
  number = {29},
  pages = {7411--7418},
  publisher = {American Chemical Society},
  doi = {10.1021/acs.jpclett.4c01500},
  urldate = {2026-05-11}
}

@article{Photoprocessing1,
  title = {Photo-Processing of Astro-{{PAHs}}},
  author = {Joblin, C and Wenzel, G and Rodriguez Castillo, S and Simon, A and Sabbah, H and Bonnamy, A and Toublanc, D and Mulas, G and Ji, M and Giuliani, A and Nahon, L},
  year = 2020,
  month = jan,
  journal = {Journal of Physics: Conference Series},
  volume = {1412},
  number = {6},
  pages = {062002},
  publisher = {IOP Publishing},
  issn = {1742-6596},
  doi = {10.1088/1742-6596/1412/6/062002},
  urldate = {2026-05-12},
  langid = {english}
}

@article{Photoprocessing2,
  title = {Astrochemical Relevance of {{VUV}} Ionization of Large {{PAH}} Cations},
  author = {Wenzel, G. and Joblin, C. and Giuliani, A. and Castillo, S. Rodriguez and Mulas, G. and Ji, M. and Sabbah, H. and Quiroga, S. and Pe{\~n}a, D. and Nahon, L.},
  year = 2020,
  month = sep,
  journal = {Astronomy \& Astrophysics},
  volume = {641},
  pages = {A98},
  publisher = {EDP Sciences},
  issn = {0004-6361, 1432-0746},
  doi = {10.1051/0004-6361/202038139},
  urldate = {2026-05-12},
  copyright = {\copyright{} G. Wenzel et al. 2020},
  langid = {english}
}

@article{Raj2014,
  title = {{{PAH Growth Initiated by Propargyl Addition}}: {{Mechanism Development and Computational Kinetics}}},
  author = {Raj, Abhijeet and Al Rashidi, Mariam J. and Chung, Suk Ho and Sarathy, S. Mani},
  year = 2014,
  month = apr,
  journal = {The Journal of Physical Chemistry A},
  volume = {118},
  number = {16},
  pages = {2865--2885},
  publisher = {American Chemical Society},
  doi = {10.1021/jp410704b}
}

@misc{KisielPIFORM,
  author    = {Kisiel, Z.},
  title     = {{PROSPE---Programs for Rotational Spectroscopy}},
  year      = {2001},
  publisher = {Institute of Physics, Polish Academy of Sciences},
  version   = {Last accessed: 2026-09-21},
  url       = {http://info.ifpan.edu.pl/%7Ekisiel/asym/asym.htm#piform}
}

@incollection{Anatomy,
  title = {Anatomy},
  booktitle = {Introduction to the {{Theory}} of {{Benzenoid Hydrocarbons}}},
  author = {Gutman, Ivan and Cyvin, Sven Josef},
  editor = {Gutman, Ivan and Cyvin, Sven Josef},
  year = 1989,
  pages = {19--32},
  publisher = {Springer},
  address = {Berlin, Heidelberg},
  doi = {10.1007/978-3-642-87143-6_3},
  urldate = {2026-09-22},
  isbn = {978-3-642-87143-6},
  langid = {english}
}

@incollection{Enumeration,
  title = {Enumeration},
  booktitle = {Introduction to the {{Theory}} of {{Benzenoid Hydrocarbons}}},
  author = {Gutman, Ivan and Cyvin, Sven Josef},
  editor = {Gutman, Ivan and Cyvin, Sven Josef},
  year = 1989,
  pages = {33--49},
  publisher = {Springer},
  address = {Berlin, Heidelberg},
  doi = {10.1007/978-3-642-87143-6_4},
  urldate = {2026-09-22},
  isbn = {978-3-642-87143-6},
  langid = {english}
}

@article{Becke:1988:3098,
  author = {Becke, A. D.},
  title = {Density-Functional Exchange-Energy Approximation with Correct Asymptotic Behavior},
  journal = {Physical Review A},
  year = {1988},
  volume = {38},
  number = {6},
  pages = {3098--3100},
  doi = {10.1103/PhysRevA.38.3098}
}

@article{Lee:1988:785,
  author = {Lee, Chengteh and Yang, Weitao and Parr, Robert G.},
  title = {Development of the Colle-Salvetti Correlation-Energy Formula into a Functional of the Electron Density},
  journal = {Physical Review B},
  year = {1988},
  volume = {37},
  number = {2},
  pages = {785--789},
  doi = {10.1103/PhysRevB.37.785}
}

@article{Stephens:1994:11623,
  author = {Stephens, P. J. and Devlin, F. J. and Chabalowski, C. F. and Frisch, M. J.},
  title = {Ab Initio Calculation of Vibrational Absorption and Circular Dichroism Spectra Using Density Functional Force Fields},
  journal = {The Journal of Physical Chemistry},
  year = {1994},
  volume = {98},
  number = {45},
  pages = {11623--11627},
  doi = {10.1021/j100096a001}
}

@article{Zhao:2008:1432,
  author = {Zhao, Yan and Truhlar, Donald G.},
  title = {The M06 Suite of Density Functionals for Main Group Thermochemistry, Thermochemical Kinetics, Noncovalent Interactions, Excited States, and Transition Elements},
  journal = {Theoretical Chemistry Accounts},
  year = {2008},
  volume = {120},
  number = {1--3},
  pages = {215--241},
  doi = {10.1007/s00214-007-0310-x}
}

@article{Chai:2008:084106,
  author = {Chai, Jeng-Da and Head-Gordon, Martin},
  title = {Systematic Optimization of Long-Range Corrected Hybrid Density Functionals},
  journal = {The Journal of Chemical Physics},
  year = {2008},
  volume = {128},
  number = {8},
  pages = {084106},
  doi = {10.1063/1.2834918}
}

@article{Becke:2006:221101,
  author = {Becke, Axel D. and Johnson, Erin R.},
  title = {A Simple Effective Potential for Exchange},
  journal = {The Journal of Chemical Physics},
  year = {2006},
  volume = {124},
  number = {22},
  pages = {221101},
  doi = {10.1063/1.2213970}
}

@article{Caldeweyher:2017:034112,
  author = {Caldeweyher, Eike and Bannwarth, Christoph and Grimme, Stefan},
  title = {Extension of the D3 Dispersion Coefficient Model},
  journal = {The Journal of Chemical Physics},
  year = {2017},
  volume = {147},
  number = {3},
  pages = {034112},
  doi = {10.1063/1.4993215}
}

@article{Caldeweyher:2019:154122,
  author = {Caldeweyher, Eike and Ehlert, Sebastian and Hansen, Andreas and Neugebauer, Hagen and Spicher, Sebastian and Bannwarth, Christoph and Grimme, Stefan},
  title = {A Generally Applicable Atomic-Charge Dependent London Dispersion Correction},
  journal = {The Journal of Chemical Physics},
  year = {2019},
  volume = {150},
  number = {15},
  pages = {154122},
  doi = {10.1063/1.5090222}
}

@article{Zheng:2011:295,
  author = {Zheng, Jingjing and Xu, Xuefei and Truhlar, Donald G.},
  title = {Minimally Augmented Karlsruhe Basis Sets},
  journal = {Theoretical Chemistry Accounts},
  year = {2011},
  volume = {128},
  number = {3},
  pages = {295--305},
  doi = {10.1007/s00214-010-0846-z}
}

@article{Kesharwani:2015:1701,
  author = {Kesharwani, Manoj K. and Brauer, Brina and Martin, Jan M. L.},
  title = {Frequency and Zero-Point Vibrational Energy Scale Factors for Double-Hybrid Density Functionals (and Other Selected Methods): Can Anharmonic Force Fields Be Avoided?},
  journal = {The Journal of Physical Chemistry A},
  year = {2015},
  volume = {119},
  number = {9},
  pages = {1701--1714},
  doi = {10.1021/jp508422u}
}

@article{Pinski:2015:034108,
  author = {Pinski, Peter and Riplinger, Christoph and Valeev, Edward F. and Neese, Frank},
  title = {Sparse Maps: A Systematic Infrastructure for Reduced-Scaling Electronic Structure Methods. I. An Efficient and Simple Linear Scaling Local MP2 Method That Uses an Intermediate Basis of Pair Natural Orbitals},
  journal = {The Journal of Chemical Physics},
  year = {2015},
  volume = {143},
  number = {3},
  pages = {034108},
  doi = {10.1063/1.4926879}
}

@article{Riplinger:2016:024109,
  author = {Riplinger, Christoph and Pinski, Peter and Becker, Ute and Valeev, Edward F. and Neese, Frank},
  title = {Sparse Maps: A Systematic Infrastructure for Reduced-Scaling Electronic Structure Methods. II. Linear Scaling Domain Based Pair Natural Orbital Coupled Cluster Theory},
  journal = {The Journal of Chemical Physics},
  year = {2016},
  volume = {144},
  number = {2},
  pages = {024109},
  doi = {10.1063/1.4939030}
}

@article{Dunning:1989:1007a,
  author = {Dunning, Jr., Thom H.},
  title = {Gaussian Basis Sets for Use in Correlated Molecular Calculations. I. The Atoms Boron through Neon and Hydrogen},
  journal = {The Journal of Chemical Physics},
  year = {1989},
  volume = {90},
  number = {2},
  pages = {1007--1023},
  doi = {10.1063/1.456153}
}

@article{Kendall:1992:6796,
  author = {Kendall, Rick A. and Dunning, Jr., Thom H. and Harrison, Robert J.},
  title = {Electron Affinities of the First-Row Atoms Revisited. Systematic Basis Sets and Wave Functions},
  journal = {The Journal of Chemical Physics},
  year = {1992},
  volume = {96},
  number = {9},
  pages = {6796--6806},
  doi = {10.1063/1.462569}
}

@article{Davidson:1996:514,
  author = {Davidson, Ernest R.},
  title = {Comment on ``Comment on Dunning's Correlation-Consistent Basis Sets''},
  journal = {Chemical Physics Letters},
  year = {1996},
  volume = {260},
  number = {3--4},
  pages = {514--518},
  doi = {10.1016/0009-2614(96)00917-7}
}

@article{Ishida:1977:2153,
  author = {Ishida, Kenichi and Morokuma, Keiji and Komornicki, Andrzej},
  title = {The Intrinsic Reaction Coordinate. An Ab Initio Calculation for {HNC} $\rightarrow$ {HCN} and {H}$^-$ + {CH}$_4$ $\rightarrow$ {CH}$_4$ + {H}$^-$},
  journal = {The Journal of Chemical Physics},
  year = {1977},
  volume = {66},
  number = {5},
  pages = {2153--2156},
  doi = {10.1063/1.434152}
}

@article{Weigend:2002:3175,
  author = {Weigend, Florian and Kohn, Andreas and Hattig, Christof},
  title = {Efficient Use of the Correlation Consistent Basis Sets in Resolution of the Identity MP2 Calculations},
  journal = {The Journal of Chemical Physics},
  year = {2002},
  volume = {116},
  number = {8},
  pages = {3175--3183},
  doi = {10.1063/1.1445115}
}

@article{Hatig:2005:59,
  author = {Hattig, Christof},
  title = {Optimization of Auxiliary Basis Sets for RI-MP2 and RI-CC2 Calculations: Core-Valence and Quintuple-Zeta Basis Sets for H to Ar and QZVPP Basis Sets for Li to Kr},
  journal = {Physical Chemistry Chemical Physics},
  year = {2005},
  volume = {7},
  number = {1},
  pages = {59--66},
  doi = {10.1039/B415208E}
}

@article{Peterson:2008:084102,
  author = {Peterson, Kirk A. and Adler, Thomas B. and Werner, Hans-Joachim},
  title = {Systematically Convergent Basis Sets for Explicitly Correlated Wavefunctions: The Atoms H, He, B-Ne, and Al-Ar},
  journal = {The Journal of Chemical Physics},
  year = {2008},
  volume = {128},
  number = {8},
  pages = {084102},
  doi = {10.1063/1.2831537}
}

@article{Noga:2009:1,
  author = {Noga, Jozef and Simunek, Jiri},
  title = {On the One-Particle Basis Set Relaxation in R12-Based Theories},
  journal = {Chemical Physics},
  year = {2009},
  volume = {356},
  number = {1--3},
  pages = {1--6},
  doi = {10.1016/j.chemphys.2008.10.012}
}

@article{Valeev:2004:190,
  author = {Valeev, Edward F.},
  title = {Improving on the Resolution of the Identity in Linear R12 Ab Initio Theories},
  journal = {Chemical Physics Letters},
  year = {2004},
  volume = {395},
  number = {4--6},
  pages = {190--195},
  doi = {10.1016/j.cplett.2004.07.061}
}

@article{Feller:1996:1571,
  author = {Feller, David},
  title = {The Role of Databases in Support of Computational Chemistry Calculations},
  journal = {Journal of Computational Chemistry},
  year = {1996},
  volume = {17},
  number = {13},
  pages = {1571--1586},
  doi = {10.1002/(SICI)1096-987X(199610)17:13<1571::AID-JCC9>3.0.CO;2-P}
}

@article{Schuchardt:2007:1045,
  author = {Schuchardt, Karen L. and Didier, Brett T. and Elsethagen, Todd and Sun, Lisong and Gurumoorthi, Vidhya and Chase, Jared and Li, Jun and Windus, Theresa L.},
  title = {Basis Set Exchange: A Community Database for Computational Sciences},
  journal = {Journal of Chemical Information and Modeling},
  year = {2007},
  volume = {47},
  number = {3},
  pages = {1045--1052},
  doi = {10.1021/ci600510j}
}
\end{CJK*}
\end{document}